\documentclass[a4paper,12pt]{article}
\usepackage{jheppub} % for details on the use of the package, please
\usepackage{tcilatex}
\usepackage{amsfonts}
\usepackage{amssymb}
\usepackage{multicol}
\usepackage{graphicx}
\usepackage{float}
\usepackage{caption}
\usepackage{xcolor}
\usepackage[utf8]{inputenc}
\usepackage{amsmath}
\title{Higher spin swampland conjecture for massive AdS$_{3}$ gravity}
\author[1,2]{R. Sammani, }
\author[1,2]{E.H Saidi}

\affiliation[1]{ LPHE-MS, Science Faculty, Mohammed V University in Rabat, Morocco}
\affiliation[2]{Centre of Physics and Mathematics, CPM- Morocco}
\emailAdd{rajae\_sammani@um5.ac.ma}
\emailAdd{e.saidi@um5r.ac.ma}
\abstract{ In this paper, we show that a possible version of the swampland weak gravity
conjecture for higher spin (HS) massive topological AdS$_{3}$ gravity can be
expressed in terms of mass \textsc{m}$_{\mathrm{hs}}$, charge \textsc{q}$_{%
\mathrm{hs}}$ and coupling constant ${\large g}_{\mathrm{hs}}$ of 3D gravity
coupled to higher spin fields as \textsc{m}$_{\mathrm{hs}}\leq \sqrt{%
2}$\textsc{q}$_{\mathrm{hs}}{\large g}_{\mathrm{hs}}M_{\mathrm{Pl}}$. The
higher spin charge\ is given by the $SO(1,2)$ quadratic Casimir \textsc{q}$_{%
\mathrm{hs}}^{2}=s\left( s-1\right) $ and the HS coupling constant by $%
{\large g}_{\mathrm{hs}}^{2}=2/\left( M_{\mathrm{Pl}}^{2}l_{\mathrm{AdS}%
_{3}}^{2}\right) $ while the mass expressed like $\left( l_{\mathrm{AdS}_{3}}%
\text{\textsc{m}}_{\mathrm{hs}}\right) ^{2}$ is defined as $\left( 1+\mathrm{%
\mu }l_{\mathrm{AdS}_{3}}\right) ^{2}s\left( s-1\right) +[1-\left( \mathrm{%
\mu }l_{\mathrm{AdS}_{3}}\right) ^{2}\left( s-1\right) ].$
}


\begin{document}
\notoc
\maketitle
\flushbottom
\newpage
\tableofcontents
\section{Introduction}

\label{sec:intro} With the purpose of investigating and deriving possible
swampland constraints \cite{swp1}-\cite{val} on topological gravity \cite%
{alvarez, holo} coupled to higher spin massive fields, we consider the three
dimensional AdS action with a negative cosmological constant in addition to
a gravitational Chern-Simons (CS) term to build up the higher spin
topological massive gravity (HSTMG) theory \cite{tmg1}-\cite{hstmg2}.
Particularly, we are interested in the higher spin Ba\~{n}%
ados-Teitelboim-Zanelli (BTZ) black hole solution \cite{BTZ, BTZ1} and its
discharge.

This study is driven by several motives, mainly the non-supersymmetric AdS
\cite{nonsusy,nonsysy2} conjecture and the weak gravity constraint \cite%
{wgc,raja,charkaoui}. In fact, it was stipulated that non supersymmetric AdS
spaces---as well as locally lookalike AdS geometries--- are at best
metastable; they manifest a non perturbative instability and will ultimately
decay \textrm{\cite{palti, val}}. And since the BTZ black hole is locally
isometric to AdS$_{3},$ it should also a priori exhibit a similar
instability \cite{alvarez}.

Actually, black holes in AdS spaces are of two types \cite{wgc3}: we either
have large black holes in equilibrium with their thermal bath, or small
unstable black holes in need of discharging by radiating away their charge.
This aligns with the weak gravity conjecture, which requires the emission of
a super-extremal particle with a constraint on its mass to charge ratio.%
\textrm{\ }However, it was argued in \cite{wgc3} that this mild version of
the WGC, demanding a single super-extremal state, is not sufficient and a
lattice refinement of the constraint is more suitable\textrm{. }To insure
the decay of BTZ black holes in AdS$_{3}$, one must guarantee that the
emitted particles reach the AdS$_{3}$ boundary and don't bounce back to form
a self-interacting particle condensate that could eventually become
sub-extremal \cite{wgc3}. Indeed, the boundary conditions on the AdS$_{3}$\
cylinder can box the discharged particles enabling them to interact in a
sub-extremal cloud of emitted particles. Therefore, one must require instead
a stronger version of the weak gravity conjecture, a lattice WGC, where in
each charge sector, there should be a super-extremal state \cite{wgc3}.%
\newline
For unstable BTZ black holes in HSTMG, and in order to comply the
super-extremality constraint of the WGC, there must be a set\ of emitted
super-extremal particles that ought to be charged and massive higher spin
particles. Now, is it possible to formulate such WGC constraint for higher
spin topological massive gravity to regulate the discharge of unstable
higher spin BTZ black holes?

To the best of our knowledge, this inquiry was never investigated in
Literature. The WGC constraint was only established for a disjointed
setting, where the gravitational and gauge sectors are separated by
considering 3D gravity in addition to a U(1) gauge field \cite{wgc3, wwgc3}
but never for massive AdS$_{3}$ gravity in CS formulation coupled to higher
spin fields.

In this paper, we intend to fill in this gap by first reviewing known
results on the D-dimensional black holes and their WGC constraints to
formulate our hypothesis about the expected super-extremality bound for the
HS-BTZ black hole (\autoref{sec:2}). Then, we construct the mass and charge
operators to build the higher spin states (\autoref{sec:3}). Once we have
all the tools needed, we derive the swampland constraint for higher spin BTZ
black holes and compute the tower of super-extremal higher spin states (%
\autoref{sec:4}). And before concluding, we discuss the relevance of our
findings with regard to recent progress in the swampland program Literature (%
\autoref{sec:5}).

\section{Weak gravity conjecture in D $\geq 4$}

\label{sec:2} \qquad This section aims to motivate a Swampland conjecture
for higher spin massive AdS$_{3}$ gravity to regulate the discharge of
higher spin BTZ black holes using commonly accepted arguments. To pave the
way for this 3D Swampland constraint, we intend to align it with the weak
gravity conjecture (WGC) governing the decays of black holes in space
dimensions $D>3$ \cite{wgc}. This bridging between HS- AdS$_{3}$ models and
effective gauge theories coupled to D-gravity (EFF$_{{\small D}}$) is
sustained by several facts and features; in particular:

$\left( \mathbf{A}\right) $ the existence of a Chern-Simons (CS) gauge
formulation of higher spin AdS$_{3}$ gravity \cite{AT,W}. In this formalism,
one uses standard gauge fields valued in the Lie algebra of the CS gauge
symmetry $\mathcal{G}_{\mathrm{hs}}\times \mathcal{\tilde{G}}_{\mathrm{hs}};$
this, as we will see, permits to replicate the construction of certain
constraints regulating the decay of D-black holes for unstable HS-BTZ black
holes. For the remainder of this investigation, we focus on higher spin BTZ
black hole solutions in the CS formulation with rank 2 gauge symmetries
namely: $\left( \mathbf{1}\right) $ the higher spin SL(3,$\mathbb{R}$) model
\cite{spin3} having two spins $s=2,3$ as a representative of the HS theory
with SL(N,$\mathbb{R}$) family \cite{slN}. $\left( \mathbf{2}\right) $ The
higher spin model with SO$\left( 2,3\right) $ group having also two spins $%
s=2,4$ as a representative of the HS ortho-symplectic families with gauge
symmetries given by the real split forms of B$_{N}$, C$_{N}$ and D$_{N}$ Lie
groups \cite{rajae}. And $\left( \mathbf{3}\right) $ the exceptional G$_{2}$
higher spin model \cite{trunc} with spin spectrum given by $s=2,6$. As well,
this G$_{2}$ can be viewed as a representative of the exceptional family of
finite dimensional Lie algebras. Useful characteristic properties of these
HS topological gauge models are as follows%
\begin{equation}
\begin{tabular}{c|c|cc}
$\ $symmetry $\mathcal{G}_{hs}$ \  & spin set $J_{\mathcal{G}}$ & generators
& $\dim \mathcal{G}_{hs}$ \\ \hline\hline
SL(3,$\mathbb{R}$) & $\ \ \ 2,$ $3$ \ \  & $\ \ \ W_{m_{1}}^{(1)}\oplus
W_{m_{{\small 2}}}^{({\small 2})}$ \ \ \  & $3+5$ \\
SO$\left( 2,3\right) $ & $\ \ \ 2,$ $4$ \ \  & $\ \ \ W_{m_{1}}^{(1)}\oplus
W_{m_{{\small 3}}}^{({\small 3})}$ \ \ \  & $3+7$ \\
G$_{2}$ & $2,$ $6$ & $W_{m_{1}}^{(1)}\oplus W_{m_{{\small 5}}}^{({\small 5}%
)} $ & $3+11$ \\ \hline\hline
\end{tabular}
\label{1}
\end{equation}%
with label $m_{j}$ taking integral values as $-j\leq m_{j}\leq j.$ The $%
W_{m_{j}}^{(j)}$ are the generators of the spin $s_{j}=j+1;$ they form an
isospin j representation of the principal sl(2;$\mathbb{R}$) partitioning
the $\mathcal{G}_{hs}$ generators as exhibited by the two last columns of
the above table.

The second feature supporting the EFT$_{{\small D}}$-HS AdS$_{3}$ cross over
is $\left( \mathbf{B}\right) $ the AdS$_{3}$/CFT$_{2}$ correspondence \cite%
{BH} allowing to relate topological aspects of HS- AdS$_{3}$ gravity such as
Wilson lines with conformal highest weight representations and conformal
observables; which will be fundamental for the computation of the HS
Swampland constraint. And lastly, the possibility to $\left( \mathbf{C}%
\right) $ realise both the masses and the charges required by the WGC in
terms of the quantum numbers of the CS gauge symmetry, as well as the
coupling constants and the Planck mass M$_{\mathrm{Pl}}$. The EFT$_{{\small D%
}}$-HS AdS$_{3}$ crossing is therefore based on matching the D- dimensional
WGC ingredients with those of 3D higher spin gravity as follows:
\begin{equation}
\begin{tabular}{|c|c|}
\hline
$D>3$ & $D=3$ \\ \hline\hline
D-gravity $+$ U(1) charged matter & 3D gravity coupled to higher spin fields
\\ \hline
Effective field models & Higher spin Chern-Simons formulation \\ \hline
Electrically charged Black holes & Higher spin BTZ black holes \\ \hline
Electric charge q$_{e}$ & Spin charge \textsc{q}$_{\mathrm{hs}}$ \\ \hline
Mass $\boldsymbol{m}$ & Mass of HS particles \textsc{m}$_{\mathrm{hs}}$ \\
\hline
\end{tabular}%
\end{equation}%
\begin{equation*}
\end{equation*}%
We show throughout this paper that the emitted particle states $\left\vert
s;\{\lambda \}\right\rangle $ of the 3D BTZ black hole carry, in addition to
the higher spins s, masses \textsc{m}$_{\mathrm{hs}}$ and charges \textsc{q}$%
_{\mathrm{hs}}$ which are functions of s. These states will be denoted below
like
\begin{equation}
|s;\text{\textsc{m}}_{\mathrm{hs}},\text{\textsc{q}}_{\mathrm{hs}}>
\end{equation}%
where the masses \textsc{m}$_{\mathrm{hs}},$ and the charges \textsc{q}$_{%
\mathrm{hs}}$ are eigenvalues of some function of commuting observables $%
\mathcal{O}_{i}$ of the gauge theory with symmetry group $\mathcal{G}_{%
\mathrm{hs}}\times \mathcal{\tilde{G}}_{\mathrm{hs}}$. Candidates for these $%
\mathcal{O}_{i}$s are given by the Cartan charge operators of $\mathcal{G}_{%
\mathrm{hs}}\times \mathcal{\tilde{G}}_{\mathrm{hs}}$ and their Casimirs.%
\newline
In this regard, we restrict to the principal SL(2,$\mathbb{R}$) symmetry
observables within the gauge symmetry $\mathcal{G}_{\mathrm{hs}}$; they are
given by the Cartan charge $L_{0}$ and the quadratic Casimir $\mathcal{C}%
_{2} $ with the following commutation relations%
\begin{eqnarray}
\lbrack L_{n},L_{m}] &=&\left( m-n\right) L_{n+m}  \label{LL} \\
\mathcal{C}_{2} &=&L_{0}^{2}-L_{0}-L_{+}L_{-}  \label{C2}
\end{eqnarray}%
Particularly, we\ are interested in the mass \textsc{\^{M}}$_{\mathrm{hs}}$
and the charge \textsc{\^{Q}}$_{\mathrm{hs}}$\ operators with spectrums as
follows%
\begin{equation}
\begin{tabular}{lll}
\textsc{m}$_{\mathrm{hs}}$ & :$=$ & {\small spect}$\left( \text{\textsc{\^{M}%
}}_{\mathrm{hs}}\right) $ \\
$\text{\textsc{q}}_{\mathrm{hs}}$ & :$=$ & {\small spect}$\left( \text{%
\textsc{\^{Q}}}_{\mathrm{hs}}\right) $%
\end{tabular}%
\end{equation}%
They can be expanded in terms of the commuting Cartan charge $L_{0}$ and the
Casimir $\mathcal{C}_{2}$ of the principal SL(2,$\mathbb{R}$) symmetry like%
\begin{eqnarray}
\text{\textsc{\^{M}}}_{\mathrm{hs}}^{2} &=&\mathfrak{m}_{0}L_{0}+\mathfrak{m}%
_{2}\mathcal{C}_{2}  \label{M} \\
\text{\textsc{\^{Q}}}_{\mathrm{hs}}^{2} &=&\mathfrak{q}_{0}L_{0}+\mathfrak{q}%
_{2}\mathcal{C}_{2}  \label{Q2}
\end{eqnarray}%
with some positive $\mathfrak{m}_{i}$ and $\mathfrak{q}_{i}$ to be
determined later on. The mass \textsc{\^{M}}$_{\mathrm{hs}}$ and charge
\textsc{\^{Q}}$_{\mathrm{hs}}$ observable operators act on the higher spin-s
particle $|s;$\textsc{m}$_{\mathrm{hs}},$\textsc{q}$_{\mathrm{hs}}>$ states
as%
\begin{equation}
\begin{tabular}{lll}
$\text{\textsc{\^{M}}}_{\mathrm{hs}}|s;$\textsc{m}$_{\mathrm{hs}},$\textsc{q}%
$_{\mathrm{hs}}>$ & $=$ & \textsc{m}$_{\mathrm{hs}}|s;$\textsc{m}$_{\mathrm{%
hs}},$\textsc{q}$_{\mathrm{hs}}>$ \\
$\text{\textsc{\^{Q}}}_{\mathrm{hs}}|s;$\textsc{m}$_{\mathrm{hs}},$\textsc{q}%
$_{\mathrm{hs}}>$ & $=$ & $\text{\textsc{q}}_{\mathrm{hs}}|s;$\textsc{m}$_{%
\mathrm{hs}},$\textsc{q}$_{\mathrm{hs}}>$%
\end{tabular}%
\end{equation}%
Before proceeding any further, we pause to carefully examine and comment on
the structure of the operators (\ref{M}-\ref{Q2}) by leveraging well-known
principles from the AdS$_{\mathrm{3}}$/CFT$_{\mathrm{2}}$\ correspondence
\cite{Cab,sahraoui} and SL$_{\mathrm{2}}$\ isospin representation framework:

\begin{description}
\item[$\left( \mathbf{1}\right) $] for the \^{M}$_{\mathrm{hs}}^{2}$
expansion (\ref{M}), the block term generated by $L_{0}$ can be attributed
to the CFT$_{2}$\ relationship $m\sim h+\bar{h}$\ with the eigenvalue
equations $L_{0}\left\vert h\right\rangle =h\left\vert h\right\rangle $ and $%
\bar{L}_{0}\left\vert \bar{h}\right\rangle =\bar{h}\left\vert \bar{h}%
\right\rangle .$\ Regarding the block term generated by $C_{2},$ it can be
motivated by the Sugawara construction of the conformal energy momentum
tensor from the affine SL$_{2}$ Kac-Moody current \cite{Goddard}.

\item[$\left( \mathbf{2}\right) $] As for the expansion of the operator \^{Q}%
$_{\mathrm{hs}}$ in (\ref{Q2}), it can be restricted to the Casimir block \^{%
Q}$_{\mathrm{hs}}^{2}=q_{2}C_{2}$ with some $q_{2}>0.$ This is because the
Casimir operator $C_{2}$ captures information on the SL$_{\mathrm{2}}$\
isospin $\Delta $\ while the charge operator $L_{0}$ (thought of as $J_{z}$%
)\ captures data on the isospin projection $\Delta _{z}.$
\end{description}

Taking all of the aforementioned into account, one might speculate that the
mass \textsc{\^{M}}$_{\mathrm{hs}}$ and the charge \textsc{\^{Q}}$_{\mathrm{%
hs}}$\ operators are indeed linked to each other like%
\begin{equation}
\text{\textsc{\^{M}}}_{\mathrm{hs}}^{2}=\mathfrak{m}_{0}L_{0}+\frac{%
\mathfrak{m}_{2}}{\mathfrak{q}_{2}}\text{\textsc{\^{Q}}}_{\mathrm{hs}}^{2}
\end{equation}%
Such property justifies the interest in the search for a Swampland
conjecture for higher spin gravitational models. Moreover, by acting on the
quantum states $\left\vert \Delta ,N\right\rangle $ of (unitary)
representations $\mathcal{R}_{\Delta }^{\pm }$ of the SL(2,$\mathbb{R}$)
symmetry group with both sides of the above equation, we get the following
mass relation%
\begin{equation}
\begin{tabular}{lllll}
$\mathcal{R}_{\Delta }^{+}$ & $:$ & $\text{\textsc{m}}_{\Delta ,N_{+}}^{2}$
& $=$ & $+\mathfrak{m}_{0}\left( \Delta +N\right) +\frac{\mathfrak{m}_{2}}{%
\mathfrak{q}_{2}}\Delta \left( \Delta -1\right) $ \\
$\mathcal{R}_{\Delta }^{-}$ & $:$ & $\text{\textsc{m}}_{\Delta ,N_{-}}^{2}$
& $=$ & $-\mathfrak{m}_{0}\left( \Delta +N\right) +\frac{\mathfrak{m}_{2}}{%
\mathfrak{q}_{2}}\Delta \left( \Delta -1\right) $%
\end{tabular}
\label{rep}
\end{equation}%
\newline
The structure of the representations $\mathcal{R}_{\Delta }^{\pm }$ and the
properties of the states $\left\vert \Delta _{\pm },N_{\pm }\right\rangle $
will be thoroughly investigated in \emph{subsection 3.1. }Meanwhile notice
that the set of HS quantum states $|s;$\textsc{m}$_{\mathrm{hs}},$\textsc{q}$%
_{\mathrm{hs}}>$ has a group theoretic basis; they can be perceived as the $%
\left\vert \Delta ,N\right\rangle $ of the SL$_{2}$ representation group
theory which will be proven to be accurate.

Returning to the HS Swampland conjecture issue, we seek to show that the
decay of small HS- BTZ black holes in AdS$_{3}$ gravity is accompanied by
the emission of super-extremal higher spin-s states $|s;\text{\textsc{m}}_{%
\mathrm{hs}},\text{\textsc{q}}_{\mathrm{hs}}>$ with spin dependent masses
\textsc{m}$_{\mathrm{hs}}$ and charges \textsc{q}$_{\mathrm{hs}}$
constrained as follows%
\begin{equation}
\text{\textsc{m}}_{\mathrm{hs}}\leq \sqrt{2}\text{\textsc{q}}_{\mathrm{hs}}%
{\large g}_{\mathrm{hs}}M_{\mathrm{Pl}}  \label{2}
\end{equation}%
with ${\large g}_{\mathrm{hs}}$ standing for the higher spin coupling
constant to be determined later. Below, we refer to (\ref{2}) as the
Swampland higher spin conjecture (\textbf{HSC}) for massive AdS$_{3}$
gravity. At first impression, one might wonder about the interpretation of
such constraint and whether this inequality is a true swampland conjecture.
However by way of construction, the swampland HSC will prove to be a version
of the WGC that regulates the discharge of higher spin BTZ back hole
solutions of HSTMG carrying charges beyond the usual U(1) of \cite{wgc3}.
The HSC accounts for 3D black holes solutions with different backgrounds
than the ones already considered in Literature \cite{wgc3, wwgc3}, and can
be therefore perceived as a complement to the work conducted in AdS$_{%
\mathrm{3}}$\ framework regarding the derivation of the WGC.

Moreover, the condition (\ref{2}) has interesting properties shared by
unstable D-black holes. Particularly, the constraint (\ref{2}) has a quite
similar structure to the well known 4D weak gravity conjecture formulated by
the following inequality \cite{wgc}%
\begin{equation}
\boldsymbol{m}\leq \sqrt{2}qg_{_{U(1)}}M_{\mathrm{Pl}}  \label{21}
\end{equation}%
This well established constraint relation (\ref{21}) will be used as a
guiding principle for the derivation of the \textbf{HSC} (\ref{2}). To avoid
confusion between the 3D and 4D parameters, we use the following convention
notations%
\begin{equation}
\begin{tabular}{l|l|l|l}
black hole & mass & charge & coupling \\ \hline\hline
4D charged BH & $\boldsymbol{m}$ & $q$ & $g_{_{U(1)}}$ \\ \hline
3D BTZ & $\text{\textsc{m}}_{\mathrm{hs}}$ & $\text{\textsc{q}}_{\mathrm{hs}%
} $ & ${\large g}_{\mathrm{hs}}$ \\ \hline\hline
\end{tabular}%
\end{equation}
\newline
In dimensions $D\geq 4,$ the weak gravity conjecture (WGC) requires the
existence of at least one super extremal state $\left\vert \boldsymbol{m}%
,q\right\rangle $ in the particle spectrum of the effective U(1) gauge
theory coupled to D- gravity with mass $\boldsymbol{m}$ and charge $%
Q_{_{U(1)}}=qg_{_{U(1)}}$satisfying the condition \cite{palti, wgc}
\begin{equation}
q^{2}g_{_{U(1)}}^{2}\geq \frac{D-3}{D-2}\boldsymbol{m}^{2}M_{\mathrm{Pl}%
}^{2-D}  \label{2A}
\end{equation}%
where the gauge coupling constant $g_{_{U(1)}}$ scales like \textsc{mass}$%
^{2-D/2}$ and $M_{\mathrm{Pl}}$\ is the D- Planck mass. This constraint
relation puts a condition on the allowed space time dimensions as it
requires $D\geq 3;$ although the $D=3$ is a critical value. By putting D=3,
the relation (\ref{2A}) leads to a trivial condition $q^{2}g_{_{U(1)}}^{2}%
\geq 0$ with no reference whatsoever to the value of the mass $\boldsymbol{m}%
^{2}.$ Even with a reverse reasoning, if we consider instead $\boldsymbol{m}%
^{2}\leq q^{2}g_{_{U(1)}}^{2}M_{\mathrm{Pl}}^{D-2}\left( D-2\right) /\left(
D-3\right) ,$ all we learn is that $\boldsymbol{m}^{2}\leq \infty $ lacking
any information on the value of $q^{2}$.

However, to retrieve additional insights, we concentrate on the interesting
four dimensional theory ($D=4).$ The condition (\ref{2A}) reads like%
\begin{equation}
\boldsymbol{m}^{2}\leq 2q^{2}g_{_{U(1)}}^{2}M_{\mathrm{Pl}}^{2}  \label{3}
\end{equation}%
with
\begin{equation}
q=\dint\nolimits_{\mathbb{S}^{2}}\mathbf{E}_{{\small u(1)}}.d\mathbf{\sigma }
\end{equation}%
and where $\mathbf{E}_{{\small u(1)}}=-\mathbf{\nabla }V-\partial _{t}%
\mathbf{A}$\ is the usual electric field. For this U(1) abelian gauge
theory, gauge group elements $\mathcal{U}$ are given by $e^{i\boldsymbol{Q}_{%
{\small u(1)}}}$ with generator
\begin{equation}
\boldsymbol{Q}_{{\small u(1)}}=g_{{\small u(1)}}\mathfrak{Q}  \label{Q}
\end{equation}%
acting on charged quantum states like
\begin{equation}
\boldsymbol{Q}_{{\small u(1)}}\left\vert \boldsymbol{m},q\right\rangle =qg_{%
{\small u(1)}}\left\vert \boldsymbol{m},q\right\rangle \qquad ,\qquad
\boldsymbol{\hat{M}}^{2}\left\vert \boldsymbol{m},q\right\rangle =%
\boldsymbol{m}^{2}\left\vert \boldsymbol{m},q\right\rangle  \label{qm4}
\end{equation}%
where $\boldsymbol{\hat{M}}$ is the mass operator. In the upcoming section,
we construct the higher spin homologue of (\ref{qm4}) for HS-AdS$_{3}$
gravity.

\section{Higher spin particle states}

\label{sec:3} An essential key component to the derivation of the relation (%
\ref{2}), is the set of emitted super extremal particle states $|s;$\textsc{m%
}$_{\mathrm{hs}},$\textsc{q}$_{\mathrm{hs}}>.$ It is therefore crucial,
before all else, to define these states. We identify these particles as
eigenstates of some higher spin charge operator defined like $\boldsymbol{Q}%
_{\mathrm{hs}}={\large g}_{\mathrm{hs}}$\textsl{Q} analogously to the 4D
charge operator $\boldsymbol{Q}_{{\small u(1)}}=g_{{\small u(1)}}\mathfrak{Q}
$ given by eq(\ref{Q}). It acts as follows
\begin{equation}
\boldsymbol{Q}_{\mathrm{hs}}\left\vert s;\text{\textsc{m}}_{\mathrm{hs}},%
\text{\textsc{q}}_{\mathrm{hs}}\right\rangle ={\large g}_{\mathrm{hs}}\text{%
\textsc{q}}_{\mathrm{hs}}\left\vert s;\text{\textsc{m}}_{\mathrm{hs}},\text{%
\textsc{q}}_{\mathrm{hs}}\right\rangle  \label{QHS}
\end{equation}%
where ${\large g}_{\mathrm{hs}}$ is the higher spin coupling constant of the
higher spin gauge theory, it will be computed later on [ see eq(\ref{cc})].%
\newline
To manoeuvre the set of these states, we use the principal SL(2,$\mathbb{R}$%
) representations since all the rank 2 gauge symmetries $\mathcal{G}_{%
\mathrm{hs}}\times \mathcal{\tilde{G}}_{\mathrm{hs}}$ we are considering can
be obtained via the principal embedding of SL(2,$\mathbb{R}$). Therefore, we
deem it necessary to briefly recall results on the principal SL(2,$\mathbb{R}
$) subgroup of the gauge symmetry $\mathcal{G}_{\mathrm{hs}}$ and its
unitary representations.

\subsection{Unitary representations of SL(2,$\mathbb{R}$)}

SL(2,$\mathbb{R}$) is a non compact group homomorphic to the Lorentz SO(1,2)
and generated by $L_{0},$ $L_{\pm }$ with commutation relations $\left[
L_{n},L_{m}\right] =(m-n)L_{n+m}$ labelled by $n,m=0,\pm $. It has several
families of irreducible representations that can classified into two sets
\textrm{\cite{BEN,BE}}, non unitary and unitary. The latter will be the
focus of the upcoming discussion. \newline
Unitary irreducible representations (UIR) are infinite dimensional, they are
obtained by requiring the hermiticity condition $L_{n}^{\dagger }=L_{-n}$
and the positivity of the quantum states norms; i.e: $\left\Vert |\psi
>\right\Vert >0$. An interesting type of these UIRs is given by the discrete
series denoted like $\mathcal{R}_{\Delta }^{\pm }$ \textrm{\cite{BAR,BEN,BE}}%
:

\begin{description}
\item[$\left( \mathbf{1}\right) $] Discrete series $\mathcal{R}_{\Delta
}^{+} $ are generated by the quantum states $\left\vert \Delta
,N\right\rangle $ as follows
\begin{equation}
\begin{tabular}{lll}
$L_{+}\left\vert \Delta ,N\right\rangle $ & $=$ & $\sqrt{\left( N+1\right)
\left( N+2\Delta \right) }\left\vert \Delta ,N+1\right\rangle $ \\
$L_{-}\left\vert \Delta ,N\right\rangle $ & $=$ & $\sqrt{N\left( N+2\Delta
-1\right) }\left\vert \Delta ,N-1\right\rangle $ \\
$L_{0}\left\vert \Delta ,N\right\rangle $ & $=$ & $\left( N+\Delta \right)
\left\vert \Delta ,N\right\rangle $ \\
$\mathcal{C}_{2}\left\vert \Delta ,N\right\rangle $ & $=$ & $\Delta \left(
\Delta -1\right) \left\vert \Delta ,N\right\rangle $%
\end{tabular}
\label{DP}
\end{equation}%
where $\mathcal{C}_{2}$ is the SL(2,$\mathbb{R}$) quadratic Casimir $%
L_{0}^{2}-L_{0}-L_{+}L_{-}.$ From these relations, one can compute useful
quantities to draw several properties; in particular: \newline
$\left( \mathbf{i}\right) $ the norm $<\Delta ,N|L_{+}L_{-}|\Delta ,N>$
which is equal to $N\left( N+2\Delta -1\right) .$ And its homologue $<\Delta
,N|L_{-}L_{+}|\Delta ,N>$ given by $\left( N+1\right) \left( N+2\Delta
\right) .$ \newline
$\left( \mathbf{ii}\right) $ The representation $\mathcal{R}_{\Delta }^{+}$
is bounded from below indicating that $L_{-}\left\vert \Delta
,N\right\rangle =0$ and requiring therefore $N\left( N+2\Delta -1\right) =0$%
. \newline
This latter constraint can be solved for $N=0,$ and the state $\left\vert
\Delta ,0\right\rangle $ with positive definite $\Delta $ is thus a lowest
weight state obeying the following lowest weight relations%
\begin{eqnarray}
L_{-}\left\vert \Delta ,0\right\rangle &=&0  \notag \\
L_{0}\left\vert \Delta ,0\right\rangle &=&\Delta \left\vert \Delta
,0\right\rangle  \label{hw} \\
\mathcal{C}_{2}\left\vert \Delta ,0\right\rangle &=&\Delta \left( \Delta
-1\right) \left\vert \Delta ,0\right\rangle  \notag
\end{eqnarray}%
With $L_{+}$ acting on $\left\vert \Delta ,0\right\rangle $ as $%
L_{+}\left\vert \Delta ,0\right\rangle =\sqrt{2\Delta }\left\vert \Delta
,1\right\rangle $.

\item[$\left( \mathbf{2}\right) $] Discrete series $\mathcal{R}_{\Delta
}^{-} $ are also generated by the states $\left\vert \Delta ,N\right\rangle $
and can be constructed as follows%
\begin{equation}
\begin{tabular}{lll}
$L_{-}\left\vert \Delta ,N\right\rangle $ & $=$ & $-\sqrt{\left( N+1\right)
\left( N+2\Delta \right) }\left\vert \Delta ,N+1\right\rangle $ \\
$L_{+}\left\vert \Delta ,N\right\rangle $ & $=$ & $-\sqrt{N\left( N+2\Delta
-1\right) }\left\vert \Delta ,N-1\right\rangle $ \\
$L_{0}\left\vert \Delta ,N\right\rangle $ & $=$ & $-\left( N+\Delta \right)
\left\vert \Delta ,N\right\rangle $ \\
$\mathcal{C}_{2}\left\vert \Delta ,N\right\rangle $ & $=$ & $\Delta \left(
\Delta -1\right) \left\vert \Delta ,N\right\rangle $%
\end{tabular}
\label{DQ}
\end{equation}%
from which we can compute: \newline
$\left( \mathbf{i}\right) $ The norm $<\Delta ,N|L_{+}L_{-}|\Delta ,N>$
giving $\left( N+1\right) \left( N+2\Delta \right) ;$ and the homologue $%
<\Delta ,N|L_{-}L_{+}|\Delta ,N>$ given by $N\left( N+2\Delta -1\right) .$
\newline
$\left( \mathbf{ii}\right) $ Conversely to the $\mathcal{R}_{\Delta }^{+}$
representation, $\mathcal{R}_{\Delta }^{-}$ is bounded from above with the
constraint $L_{+}\left\vert \Delta ,N\right\rangle =0$ requiring $N\left(
N+2\Delta -1\right) =0$. \newline
Analogously, if we impose $N=0,$ the highest weight state $\left\vert \Delta
,0\right\rangle $ annihilated by $L_{+}$ satisfies the relations%
\begin{eqnarray}
L_{+}\left\vert \Delta ,0\right\rangle &=&0  \notag \\
L_{0}\left\vert \Delta ,0\right\rangle &=&-\Delta \left\vert \Delta
,0\right\rangle \\
\mathcal{C}_{2}\left\vert \Delta ,0\right\rangle &=&\Delta \left( \Delta
-1\right) \left\vert \Delta ,0\right\rangle  \notag
\end{eqnarray}%
with $L_{-}$ action given by $L_{-}\left\vert \Delta ,0\right\rangle =-\sqrt{%
2\Delta }\left\vert \Delta ,-1\right\rangle $.
\end{description}

Notice that the two discrete representations $\mathcal{R}_{\Delta }^{+}$ and
$\mathcal{R}_{\Delta }^{-}$ are isomorphic; the isomorphism $\iota :\mathcal{%
R}_{\Delta }^{+}$ $\rightarrow \mathcal{R}_{\Delta }^{-}$ is given by the
1:1 correspondence $\iota \left( L_{n}\right) =-L_{-n}$ as manifestly
exhibited by the relations (\ref{DP}) and (\ref{DQ}).

\subsection{Higher spin AdS$_{3}$ gravity}

Focussing on HS- BTZ black holes with rank 2 symmetries of eq(\ref{1}), the
gauge theory is described by the 3D HS gravity action $\mathcal{S}_{{\small 0%
}}^{\text{\textsc{grav}}}$ given in terms of two copies of Chern-Simons (CS)
fields $A$ and $\tilde{A}$ as follows \cite{AT, W}%
\begin{equation}
\mathcal{S}_{{\small 0}}^{\text{\textsc{grav}}}=\frac{\mathrm{k}}{4\pi }\int
tr(AdA+\frac{2}{3}A^{3}\mathbf{)-}\frac{\mathrm{\tilde{k}}}{4\pi }\int tr(%
\tilde{A}d\tilde{A}+\frac{2}{3}\tilde{A}^{3})  \label{4}
\end{equation}%
with CS level $\mathrm{\tilde{k}=k.}$ This positive integer number is
related to the AdS$_{3}$ radius and the 3D Newton coupling constant like $%
\mathrm{k}=l_{\mathrm{AdS}_{{\small 3}}}/\left( 4G_{N}\right) .$ Being a
discrete relation, this quantity can be imagined as a quantization relation
of the 3D Newton constant expressed like $G_{N}^{{\small [\mathrm{k}]}}=l_{%
\mathrm{AdS}_{{\small 3}}}/(4\mathrm{k}),$ showing in turns that $G_{N}^{%
{\small [1]}}=l_{\mathrm{AdS}_{{\small 3}}}/4.$

The conversion to the metric formulation is quite straightforward and mainly
based on expressing both the dreibein $E_{\mu }$ and the spin connection $%
\Omega _{\mu }$ in terms of the two CS gauge potentials $A_{\mu }$ and $%
\tilde{A}_{\mu }$ as follows%
\begin{equation}
\begin{tabular}{lll}
$G_{\mu \nu }$ & $=$ & $\frac{1}{2}Tr\left( E_{\mu }E_{\nu }\right) $ \\
$\Phi _{\mu _{1}...\mu _{s}}$ & $=$ & $Tr\left( E_{(\mu _{1}}...E_{\mu
_{s})}\right) $ \\
$E_{\mu }$ & $=$ & $A_{\mu }-\tilde{A}_{\mu }$ \\
$\Omega _{\mu }$ & $=$ & $A_{\mu }+\tilde{A}_{\mu }$%
\end{tabular}%
\end{equation}%
Notice also that here the 1-form gauge connections $A$ and $\tilde{A}$ as
well as the $E$ and $\Omega $ are non abelian 3D fields; they are valued in
the Lie algebra of the gauge symmetry $\mathcal{G}_{hs}\times \mathcal{%
\tilde{G}}_{hs}$ and satisfy the Grumiller-Riegler (GR) boundary conditions
\textrm{\cite{GR} }for a more general set-up. The field equations of motion
of (\ref{4}) are given by $\mathcal{F}_{\mu \nu }=0$ and $\mathcal{\tilde{F}}%
_{\mu \nu }=0$ where the $\mathcal{F}_{\mu \nu }$ and $\mathcal{\tilde{F}}%
_{\mu \nu }$ are the gauge fields strengths reading as $\partial _{\mu
}A_{\nu }-\partial _{\nu }A_{\mu }+[A_{\mu },A_{\nu }]$ and $\partial
_{\lbrack \mu }\tilde{A}_{\nu ]}+[\tilde{A}_{\mu },\tilde{A}_{\nu }].$

Because of the vanishing value of the gauge field strengths, gauge
invariants similar to the 4D electric field $\mathbf{E}_{{\small u(1)}}$ of (%
\ref{4}) and the associated electric charge $q=\int_{\mathbb{S}^{2}}\mathbf{E%
}_{{\small u(1)}}.d\mathbf{\sigma }$ are unavailable in the higher spin AdS$%
_{3}$ gravity. Instead, there are alternative gauge invariants given by $%
\left( \mathbf{i}\right) $ the Wilson loops%
\begin{equation}
\mathcal{W}_{\mathcal{R}}\left[ \mathrm{\gamma }\right] =Tr_{\mathcal{R}}%
\left[ P\exp \left( \dint\nolimits_{\mathrm{\gamma }}\boldsymbol{A}\right)
P\exp \left( \dint\nolimits_{\mathrm{\gamma }}\boldsymbol{\tilde{A}}\right) %
\right]  \label{5}
\end{equation}%
with $\mathcal{R}$ being a representation of the gauge symmetry, $\mathrm{%
\gamma }$ a loop in AdS$_{3}$, $\boldsymbol{A}$ as well as $\boldsymbol{%
\tilde{A}}$\ some gauge connections expanding as $A_{\mu }dx^{\mu }$ and $%
\tilde{A}_{\mu }dx^{\mu }.$ And $\left( \mathbf{ii}\right) $ topological
defects given by line operators constructed as \cite{castro,EHS}
\begin{equation}
\mathcal{W}_{\mathcal{R}}\left[ y_{i},y_{f}\right] =\left\langle
U_{i}\right\vert Tr_{\mathcal{R}}\left[ P\exp \left( \dint\nolimits_{\mathrm{%
\Upsilon }_{if}}\boldsymbol{A}\right) P\exp \left( \dint\nolimits_{\mathrm{%
\Upsilon }_{if}}\boldsymbol{\tilde{A}}\right) \right] \left\vert
U_{f}\right\rangle  \label{6}
\end{equation}%
where $\left( y_{i},y_{f}\right) $ are the end points of the curve $\mathrm{%
\Upsilon }_{if}$ parameterised by $y$ and where $U\left( y\right) $ is a
probe field on $\mathrm{\Upsilon }_{if}$ with boundary condition $U\left(
y_{i}\right) =U\left( y_{f}\right) =I_{id}$. As illustrations, we give the
expansion of the potential $A_{\mu }$ for the SL(3,$\mathbb{R}$) and G$_{2}$
models in the higher spin basis. For SL(3,$\mathbb{R}$), we have the
following splitting%
\begin{equation}
\begin{tabular}{cclll}
SL(3,$\mathbb{R}$) & : & $A_{\mu }$ & $=$ & $\dsum\limits_{m_{1}=-1}^{1}%
\mathcal{A}_{\mu }^{m_{1}}W_{m_{1}}^{(1)}+\dsum\limits_{m_{2}=-2}^{2}%
\mathcal{W}_{\mu }^{m_{2}}W_{m_{{\small 2}}}^{({\small 2})}$ \\
&  &  & $:=$ & $\dsum\limits_{m=-1}^{1}\mathcal{A}_{\mu
}^{m}L_{m}+\dsum\limits_{n=-2}^{2}\mathcal{W}_{\mu }^{n}W_{n}$%
\end{tabular}%
\end{equation}%
with the commutation relations \cite{compleoni}%
\begin{equation}
\begin{tabular}{lll}
$\left[ L_{i},L_{j}\right] $ & $=$ & $\left( j-i\right) L_{i+j}$ \\
$\left[ L_{i},W_{m}\right] $ & $=$ & $\left( m-2i\right) W_{i+m}$ \\
$\left[ W_{n},W_{m}\right] $ & $=$ & $\frac{1}{3}\left( n-m\right) \left(
2m^{2}+2n^{2}-mn-8\right) L_{n+m}$%
\end{tabular}
\label{cr}
\end{equation}%
Similarly for G$_{2}$, we can write%
\begin{equation}
\begin{tabular}{cclll}
G$_{2}$ & : & $A_{\mu }$ & $=$ & $\dsum\limits_{m_{1}=-1}^{1}\mathcal{A}%
_{\mu }^{m_{1}}W_{m_{1}}^{(1)}+\dsum\limits_{m_{5}=-5}^{5}\mathcal{W}_{\mu
}^{m_{5}}W_{m_{{\small 5}}}^{({\small 5})}$ \\
&  &  & $:=$ & $\dsum\limits_{m=-1}^{1}\mathcal{A}_{\mu
}^{m}L_{m}+\dsum\limits_{n=-5}^{5}\mathcal{W}_{\mu }^{n}W_{n}$%
\end{tabular}%
\end{equation}%
with%
\begin{equation}
\begin{tabular}{lll}
$\left[ L_{i},L_{j}\right] $ & $=$ & $\left( j-i\right) L_{i+j}$ \\
$\left[ L_{i},W_{m}\right] $ & $=$ & $\left( m-5i\right) W_{i+m}$ \\
$\left[ W_{n},W_{m}\right] $ & $=$ & $f_{m,n|2}^{\left( 1,5\right) }L_{n+m}$%
\end{tabular}%
\end{equation}%
where $f_{m,n|2}^{\left( 1,5\right) }$ are constant structures obtained by
solving the Jacobi identities. Notice that in the HS- basis, the 8
generators of SL(3,$\mathbb{R}$) are split into two blocks 3+5 given by: $%
\left( \mathbf{i}\right) $ the three $W_{m_{1}}^{(1)}$ with label $%
m_{1}=0,\pm 1;$ they are just the usual generators $L_{m}$ of the principal
SL(2,$\mathbb{R}$)$.$ $\left( \mathbf{ii}\right) $ The five $%
W_{m_{2}}^{(2)}\equiv W_{n}$ with index $m_{2}=n=0,\pm 1,\pm 2;$ they
generate the coset space SL(3,$\mathbb{R}$)/SL(2,$\mathbb{R}$). Quite
similar relations can be written for the 14 generators of G$_{2}$ that split
as $3+11.$

As far as these types of HS- expansions are concerned, notice the following
features depicted for the case of SL(3,$\mathbb{R}$) model: $\left( \mathbf{i%
}\right) $ The commutation relations of sl(3,$\mathbb{R}$) in the HS basis
can be presented in a condensed form as follows%
\begin{equation}
\left[ W_{m_{j}}^{(j)},W_{n_{k}}^{(k)}\right] =\dsum\limits_{r_{1}=-1}^{1}%
\mathrm{f}_{n_{k},m_{j}|1}^{{\small (j,k)}}\delta
_{m_{j}+n_{k}}^{r_{1}}W_{r_{1}}^{{\small (1)}}+\dsum\limits_{r_{2}=-2}^{2}%
\mathrm{f}_{n_{k},m_{j}|2}^{{\small (j,k)}}\delta
_{m_{j}+n_{k}}^{r_{2}}W_{r_{2}}^{{\small (2)}}
\end{equation}%
with the constant structures $\mathrm{f}_{m_{j},n_{k}|s}^{{\small (j,k)}}$
given by
\begin{equation}
\begin{tabular}{lll}
$\mathrm{f}_{n_{1},m_{1}|1}^{{\small (1,1)}}$ & $=$ & $m_{1}-n_{1}$ \\
$\mathrm{f}_{n_{2},m_{1}|2}^{{\small (1,2)}}$ & $=$ & $m_{1}-2n_{2}$ \\
$\mathrm{f}_{n_{2},m_{2}|2}^{{\small (2,2)}}$ & $=$ & $\frac{1}{3}\left(
n_{2}-m_{2}\right) \left( 2m_{2}^{2}+2n_{2}^{2}-m_{2}n_{2}-8\right) $%
\end{tabular}%
\end{equation}%
In general, we can express these commutations in a shorter form like%
\begin{equation}
\left[ W_{m_{\tau }}^{(\tau )},W_{n_{\sigma }}^{(\sigma )}\right]
=\sum_{\upsilon }\dsum\limits_{r_{\upsilon }}\mathrm{f}_{n_{\sigma },m_{\tau
}|\upsilon }^{{\small (\tau ,\sigma )}}\delta _{m_{\tau }+n_{\sigma
}}^{r_{\upsilon }}W_{r_{\upsilon }}^{{\small (\upsilon )}}
\end{equation}%
$\left( \mathbf{ii}\right) $ Higher spin theories are characterised by the
spins- s of the principal SL(2,$\mathbb{R}$) within SL(3,$\mathbb{R}$); it
is defined by the usual commutation relations (\ref{LL}) where we have set $%
L_{m}=W_{m_{1}}^{({\small 1})}$. As such, it is interesting to use the
formal decomposition
\begin{equation}
SL(3,\mathbb{R})=SL(2,\mathbb{R})\ltimes \frac{SL(3,\mathbb{R})}{SL(2,%
\mathbb{R})}
\end{equation}%
to split the gauge potentials $A_{\mu }$ and $\tilde{A}_{\mu }$ as follows%
\begin{equation}
\begin{tabular}{lll}
$A_{\mu }$ & $=$ & $A_{\mu }^{{\small sl}_{{\small 2}}}+A_{\mu }^{{\small sl}%
_{{\small 3/2}}}$ \\
$\tilde{A}_{\mu }$ & $=$ & $\tilde{A}_{\mu }^{{\small sl}_{{\small 2}}}+%
\tilde{A}_{\mu }^{{\small sl}_{{\small 3/2}}}$%
\end{tabular}%
\end{equation}

\section{Derivation of the HS Swampland conjecture}

\label{sec:4} \qquad In this section, we target the derivation of the HS
Swampland conjecture in AdS$_{3}$ (\ref{2})\ which can be articulated as in
the following statement:

\textbf{Higher spin Swampland conjecture in AdS}$_{3}$\newline
\emph{A higher spin BTZ black hole solution of 3D topologically massive
gravity with a negative cosmological constant} $\Lambda <0$ \emph{should be
able to discharge by emitting super-extremal higher spin particles with mass}
\textsc{m}$_{\mathrm{hs}}$ \emph{and charge} \textsc{q}$_{\mathrm{hs}}$
\emph{such that}
\begin{equation}
\text{\textsc{m}}_{\mathrm{hs}}\leq \mathtt{\alpha }_{3}\text{\textsc{q}}_{%
\mathrm{hs}}{\large g}_{\mathrm{hs}}M_{\mathrm{Pl}}  \label{QS}
\end{equation}%
\emph{where} ${\large g}_{\mathrm{hs}}$ \emph{is the higher spin gauge
coupling and }$\mathtt{\alpha }_{3}$ \emph{is some constant that we set as} $%
\mathtt{\alpha }_{3}=\sqrt{2}$\emph{.}

The constraint (\ref{QS}) bears a mighty resemblance to the inequality (\ref%
{3}) regulating the decay of charged 4D black holes,%
\begin{equation}
\begin{tabular}{ccccccc}
\multicolumn{3}{c}{3D HS-BTZ} & $\leftrightarrow $ & \multicolumn{3}{c}{4D
charged BH} \\
$\text{\textsc{m}}_{\mathrm{hs}}$ & $\leq $ & $\sqrt{2}\text{\textsc{q}}_{%
\mathrm{hs}}{\large g}_{\mathrm{hs}}M_{\mathrm{Pl}}$ & $\qquad
\leftrightarrow \qquad $ & $\boldsymbol{m}$ & $\leq $ & $\sqrt{2}%
qg_{_{U(1)}}M_{\mathrm{Pl}}$%
\end{tabular}
\label{34D}
\end{equation}%
but instead of the abelian U(1) parameters, we must determine the higher
spin \textsc{m}$_{\mathrm{hs}}$, \textsc{q}$_{\mathrm{hs}}$ and ${\large g}_{%
\mathrm{hs}}$ quantities for the $\mathcal{G}_{\mathrm{hs}}$ symmetry. For
this purpose, we first promote the 3D gravity theory described by the field
action $\mathcal{S}_{{\small 0}}^{\text{\textsc{grav}}}$ to a higher spin
topologically massive AdS$_{3}$ gravity \textrm{\cite{tmg1,hstmg1}} in order
for our, as of yet, massless higher spin states to acquire mass.

The pure 3D gravity is known to be topological due to the absence of local
degrees of freedom, offering simple settings that allow for tractable
studies of gravitational theories, including those coupled to higher spin
fields. This pure theory can be extended by incorporating massive degrees of
freedom, by deforming the AdS$_{3}$ action with a gravitational CS term%
\textbf{\ \cite{tmg1}-\cite{hstmg2}:}%
\begin{equation}
\mathcal{S}_{\text{\textsc{1}}}^{\text{\textsc{grav}}}=\frac{M_{\mathrm{Pl}}%
}{2\mathrm{\mu }}\int_{\mathcal{M}_{3D}}Tr\left( \Gamma d\Gamma +\frac{2}{3}%
\Gamma ^{3}\right)  \label{gama}
\end{equation}%
where $\Gamma $ is the Christoffel symbol and where $\mathrm{\mu }$ is a
massive parameter. The modified equations of motion are as follows:%
\begin{equation}
G_{\mu \nu }+\frac{1}{\mathrm{\mu }}C_{\mu \nu }=0
\end{equation}%
where the Einstein tensor is given by:%
\begin{equation}
G_{\mu \nu }=R_{\mu \nu }-\frac{1}{2}g_{\mu \nu }R-\frac{1}{l_{AdS_{3}}^{2}}%
g_{\mu \nu }
\end{equation}%
and the Cotton tensor by:%
\begin{equation}
C_{\mu \nu }=\frac{1}{2}\varepsilon _{\mu }^{\alpha \beta }\nabla _{\alpha
}R_{\beta \nu }+\left( \mu \leftrightarrow \nu \right)
\end{equation}%
The theory develops a diffeomorphism anomaly given by the difference between
the right $c_{+}$ and the left $c_{-}$ central charges as we will discuss
more thoroughly in the next section.\newline
Now, regarding the mass of the new massive mode of topologically massive
spin 2 gravity has been computed in \cite{22, 23} and is given by:%
\begin{equation}
m_{\left( 2\right) }^{2}=\left( \mathrm{\mu }+\frac{2}{l_{\mathrm{AdS}_{%
{\small 3}}}}\right) ^{2}-\frac{1}{l_{\mathrm{AdS}_{{\small 3}}}^{2}}=\frac{1%
}{l_{\mathrm{AdS}_{{\small 3}}}^{2}}\left( \mathrm{\mu }l_{\mathrm{AdS}_{%
{\small 3}}}+3\right) \left( \mathrm{\mu }l_{\mathrm{AdS}_{{\small 3}%
}}+1\right)
\end{equation}%
It can also be written as:
\begin{equation}
m_{\left( 2\right) }^{2}=\frac{\left( 2-1\right) }{l_{\mathrm{AdS}_{{\small 3%
}}}^{2}}\left( \left( 2-1\right) \mathrm{\mu }l_{\mathrm{AdS}_{{\small 3}%
}}+\left( 2+1\right) \right) \left( \mathrm{\mu }l_{\mathrm{AdS}_{{\small 3}%
}}+1\right)
\end{equation}%
A possible generalisation of the s=2 formula for higher spin super extremal
states \textsc{m}$_{\mathrm{hs}}$ is therefore a function of the parameter $%
\mathrm{\mu }$ and the conformal spin s such that \textsc{m}$_{\mathrm{hs}}=$%
\textsc{m}$\left( \mathrm{s},\mathrm{\mu }\right) $. Following the
conjecture of \textrm{\cite{alvarez, 22, 23},} \textsc{m}$_{\mathrm{hs}}$
can be formulated as
\begin{equation}
\text{\textsc{m}}_{\mathrm{hs}}^{2}=\frac{1+\mathrm{\mu }l_{\mathrm{AdS}_{3}}%
}{l_{\mathrm{AdS}_{3}}^{2}}\left( s-1\right) \left[ \left( s-1\right)
\mathrm{\mu }l_{\mathrm{AdS}_{3}}+\left( s+1\right) \right]  \label{ms}
\end{equation}
This is a remarkable relation that can be put into a covariant form using
observables of the principal SL(2,$\mathbb{R}$) symmetry of the higher spin
theory. In fact, by putting $M_{\mathrm{AdS}_{3}}=1/l_{\mathrm{AdS}_{3}}$
into the above \textsc{m}$_{\mathrm{hs}}^{2}$ relation and after rearranging
the terms, we end up with the distinguishable expression%
\begin{equation}
\text{\textsc{m}}_{\mathrm{hs}}^{2}=\left( M_{\mathrm{AdS}_{3}}+\mathrm{\mu }%
\right) ^{2}s\left( s-1\right) +\left( M_{\mathrm{AdS}_{3}}^{2}-\mathrm{\mu }%
^{2}\right) \left( s-1\right)  \label{sm}
\end{equation}%
more thoroughly investigated below. The relationship between this
conjectured mass formula and those \textsc{m}$_{\Delta ,N_{+}}^{2}$ and
\textsc{m}$_{\Delta ,N_{-}}^{2}$ given by (\ref{rep}), associated with the
two unitary representations $\mathcal{R}_{\Delta }^{+}$ and $\mathcal{R}%
_{\Delta }^{-}$, will be commented\textrm{\ }in subsection 4.2. Before that,
let us see how the swampland constraint relation can be derived from (\ref%
{sm}).

\subsection{From eq(\protect\ref{sm}) towards eq(\protect\ref{QS})}

Using the relation $s=1+j,$ linking the values of the conformal spins- s of
the higher spin AdS$_{3}$ gravity to the isospin j representation weights of
SL(2,$\mathbb{R}$), the above conjectured mass relation \textsc{m}$_{\mathrm{%
hs}}^{2}$ becomes%
\begin{eqnarray}
\text{\textsc{m}}_{\mathrm{hs}}^{2} &=&\left( M_{\mathrm{AdS}_{3}}+\mathrm{%
\mu }\right) ^{2}j\left( j+1\right) +\left( M_{\mathrm{AdS}_{3}}^{2}-\mathrm{%
\mu }^{2}\right) j  \notag \\
&=&\left( \frac{1+\mathrm{\mu }l_{\mathrm{AdS}_{3}}}{l_{\mathrm{AdS}_{3}}}%
\right) ^{2}j\left( j+1\right) +\left( \frac{1-\mathrm{\mu }^{2}l_{\mathrm{%
AdS}_{3}}^{2}}{l_{\mathrm{AdS}_{3}}^{2}}\right) j  \label{M2}
\end{eqnarray}%
exhibiting two well known quantum numbers of SL(2,$\mathbb{R}$)
representations namely the second Casimir $j\left( j+1\right) $ and the
Cartan charge $j$ of highest weight (HW) state. By framing (\ref{ms}) in the
form (\ref{M2}), the dependence on the second Casimir comes as no surprise
especially in holographic contexts like ours. In AdS/CFT, the mass spectrum
of states is encoded in the eigenvalues of the Casimir operator. In fact,
the action of the quadratic Casimir is related to the mass of the quantum
states as follows \cite{mass}-\cite{mass4}:
\begin{equation}
m^{2}l_{AdS}^{2}=C_{2}
\end{equation}%
exactly as given by the first half of (\ref{M2}) namely:%
\begin{equation}
\text{\textsc{\^{M}}}_{\mathrm{hs}}^{2}=\mathfrak{m}_{2}\mathcal{C}_{2}
\end{equation}%
However, a state in the spectrum is uniquely identified once considering
both the Casimir and its Cartan charge $(j(j+1),j).$ For example, states in
the same HW representation share the same Casimir value, so to distinguish
between individual states within that representation, we must take into
account, in addition to the Casimir, the Cartan charge operator $L_{0}$ as
given in (\ref{M}):%
\begin{equation}
\text{\textsc{\^{M}}}_{\mathrm{hs}}^{2}=\mathfrak{m}_{0}L_{0}+\mathfrak{m}%
_{2}\mathcal{C}_{2}
\end{equation}%
Then, we use the conjecture to identify and established a formula of the
coefficients $\mathfrak{m}_{0}$ and $\mathfrak{m}_{2}$\ in terms of the
parameters of the HSTMG theory at hand, we have:%
\begin{eqnarray}
\mathfrak{m}_{2} &=&\left( \frac{1+\mathrm{\mu }l_{\mathrm{AdS}_{3}}}{l_{%
\mathrm{AdS}_{3}}}\right) ^{2} \\
\mathfrak{m}_{0} &=&\left( \frac{1-\mathrm{\mu }^{2}l_{\mathrm{AdS}_{3}}^{2}%
}{l_{\mathrm{AdS}_{3}}^{2}}\right)
\end{eqnarray}%
giving thus (\ref{M2}).

Moreover, since $j\geq 1$ due to the condition $s\geq 2,$ we have the
property $j\left( j+1\right) >j$ implying that the dominant term in the
\textsc{m}$_{\mathrm{hs}}^{2}$ formula is given by the block term $\left( M_{%
\mathrm{AdS}_{3}}+\mathrm{\mu }\right) ^{2}j\left( j+1\right) .$
Furthermore, we can note two additional valuable features:

$\left( \mathbf{i}\right) $ In the region of the parameter space of the
higher spin theory where $M_{\mathrm{AdS}_{3}}^{2}-\mathrm{\mu }^{2}$ is
negative definite ( i.e: $1-\mathrm{\mu }^{2}l_{\mathrm{AdS}_{3}}^{2}<0);$
we have
\begin{equation}
\mathrm{\mu }^{2}>M_{\mathrm{AdS}_{3}}^{2}\qquad \Leftrightarrow \qquad
\mathrm{\mu }^{2}>\frac{1}{l_{\mathrm{AdS}_{3}}^{2}}
\end{equation}%
and then the mass formula (\ref{M2}) induces the following inequality%
\begin{equation}
\text{\textsc{m}}_{\mathrm{hs}}^{2}<\left( M_{\mathrm{AdS}_{3}}+\mathrm{\mu }%
\right) ^{2}j\left( j+1\right) \qquad \Leftrightarrow \qquad \text{\textsc{m}%
}_{\mathrm{hs}}^{2}<\left( \frac{1+\mathrm{\mu }l_{\mathrm{AdS}_{3}}}{l_{%
\mathrm{AdS}_{3}}}\right) ^{2}j\left( j+1\right)  \label{ct}
\end{equation}%
which corresponds precisely to (\ref{QS}); thus offering a natural candidate
for the swampland conjecture regarding HS topological AdS$_{3}$ massive
gravity.

$\left( \mathbf{ii}\right) $ For the critical value $\mathrm{\mu }^{2}=%
\mathrm{\mu }_{c}^{2}=M_{\mathrm{AdS}_{3}}^{2}$, the block term $\left( M_{%
\mathrm{AdS}_{3}}^{2}-\mathrm{\mu }_{c}^{2}\right) j$ in (\ref{M2})\textrm{\
}vanishes; and the mass formula \textsc{m}$_{\mathrm{hs}}^{2}$ in (\ref{M2})
is equal to $($\textsc{m}$_{\mathrm{hs}}^{2})_{c}=$ $4M_{\mathrm{AdS}%
_{3}}^{2}j\left( j+1\right) .$\newline
So, using $\mathrm{\mu }^{2}\simeq M_{\mathrm{AdS}_{3}}^{2}+\delta \mathrm{%
\mu }^{2}$ with positive $\delta \mathrm{\mu }^{2},$ eq(\ref{M2}) becomes%
\begin{equation}
\text{\textsc{m}}_{\mathrm{hs}}^{2}=4M_{\mathrm{AdS}_{3}}^{2}j\left(
j+1\right) -\left( \delta \mathrm{\mu }^{2}\right) j
\end{equation}%
thus leading to the inequality%
\begin{equation}
\text{\textsc{m}}_{\mathrm{hs}}^{2}\leq 4M_{\mathrm{AdS}_{3}}^{2}j\left(
j+1\right) \qquad \Leftrightarrow \qquad \text{\textsc{m}}_{\mathrm{hs}%
}^{2}\leq \frac{4}{l_{\mathrm{AdS}_{3}}^{2}}j\left( j+1\right)
\end{equation}%
In comparison with (\ref{QS}) stipulating \textsc{m}$_{\mathrm{hs}}^{2}\leq 2
$\textsc{q}$_{\mathrm{hs}}^{2}{\large g}_{\mathrm{hs}}^{2}M_{\mathrm{Pl}%
}^{2},$ one can deduce the expressions of both the charge \textsc{q}$_{%
\mathrm{hs}}$ and the coupling constant ${\large g}_{\mathrm{hs}};$ they are
given by%
\begin{eqnarray}
\text{\textsc{q}}_{\mathrm{hs}}^{2} &=&j\left( j+1\right) =s\left(
s-1\right)   \label{jj} \\
{\large g}_{\mathrm{hs}}^{2} &=&\frac{2M_{\mathrm{AdS}_{3}}^{2}}{M_{\mathrm{%
Pl}}^{2}}=\frac{2}{M_{\mathrm{Pl}}^{2}l_{\mathrm{AdS}_{3}}^{2}}  \label{cc}
\end{eqnarray}%
with higher spin $s=1+j.$ The expression of the coupling constant (\ref{cc})
can be presented otherwise by using the CS level relation $\mathrm{k}=l_{%
\mathrm{AdS}_{{\small 3}}}/(4G_{N}),$ which gives
\begin{equation}
{\large g}_{\mathrm{hs}}^{2}=\frac{1}{8\mathrm{k}^{2}G_{N}^{2}M_{\mathrm{Pl}%
}^{2}}
\end{equation}%
where the dependence on the Chern-Simons coupling $\mathrm{k}$, the Newton
constant $G_{N}$ as well as Planck mass $M_{\mathrm{Pl}}$ is exhibited. As
these constant are interconnected, we can further unclutter the expression
by using the relation $M_{\mathrm{Pl}}G_{N}=1/\left( 8\pi \right) $ to
showcase that ${\large g}_{\mathrm{hs}}$ is merely the inverse of the
Chern-Simons \textrm{k}:
\begin{equation}
{\large g}_{\mathrm{hs}}^{2}=\left( 8\pi ^{2}\right) /\mathrm{k}^{2}
\label{cc2}
\end{equation}

To justify interpreting $j(j+1)=s(s-1)$ as a higher spin charge, we draw
connections to results from the Literature on charged black holes in higher
dimensions, including comparisons with: \textbf{(i)} the extremality
constraints for Kerr-Newman black hole in D-dimensions, and \textbf{(ii)}
the extremality constraint on HS-BTZ entropy.

We begin by recalling that the mass spectrum of charged states in
D-dimensional effective field theories coupled to gravity (D\TEXTsymbol{>}3)
can be categorised into two distinct regimes based on their mass to charge
ratios, as stipulated by the weak gravity conjecture \textrm{\cite{RWGC}.}
These regimes are referred to as sub-extremal and super-extremal, as
described below:

\textbf{(a)- \emph{sub-extremal regime}:}\newline
This phase consists of massive charged states $\left \vert M_{{\small BH}%
},Q_{{\small BH}}\right \rangle $ whose mass is bounded from below like $M_{%
{\small BH}}^{2}\geq \frac{g^{2}}{\sqrt{G}}Q_{{\small BH}}^{2}$ where $g$ is
the gauge coupling of the charge symmetry and $G$ is the "Newton" constant.
These states correspond to black holes solutions and are to referred to as
sub-extremal. In the particular instance where $\left( M_{{\small BH}%
}^{2}\right) _{ext}$ is equal to $\frac{g^{2}}{\sqrt{G}}\left( Q_{{\small BH}%
}^{2}\right) _{ext}$, the charged black hole is said to be extremal; its
mass is \textrm{given by} the charge, up to a multiplicative constant.

\textbf{(b)- \emph{super-extremal regime}:}\newline
This phase is composed of quantum particle states $\left \vert
m,Q\right
\rangle $ with mass bounded from above like $m^{2}\leq \frac{g^{2}%
}{\sqrt{G}}Q^{2}.$ In this regime, the masses of the states satisfy $m<M_{%
{\small BH}}$; they describe super-extremal particles emitted during the
discharge of the black hole. The kinematical properties of these states are
governed by two main conservation laws: the total energy-momentum
conservation \textrm{inducing }the mass inequality $\sum_{i}m_{i}\leq M_{BH}$
and the charge conservation given by the equation $Q_{BH}=\sum_{i}Q_{i}$.
Following \textrm{\cite{MI}}, the presence of super-extremal particles
ensures the decay of the black hole by maintaining $\frac{Q}{m}\geq (\frac{%
Q_{{\small BH}}}{M_{{\small BH}}})_{ext}\sim O\left( 1\right) $. Notice that
the existence of super-extremal states is necessary for the consistency of
the quantum gravity theory. It is also worth noting that particle states
with the particular mass $m^{2}=\frac{g^{2}}{\sqrt{G}}Q^{2}$ can be viewed
as BPS-like states.

In our setting, we similarly identify two distinct regimes \textbf{(a)} and
\textbf{(b)}. First, the sub-extremal higher spin BTZ black hole correspond%
\textrm{ing} to a state $\left\vert M_{{\small BH}},Q_{{\small BH}%
}\right\rangle $ in regime-\textbf{(a),} characterised by the inequality $M_{%
{\small BH}}^{2}\geq \frac{g^{2}}{\sqrt{G}}Q_{{\small BH}}^{2}.$ The second
regime-\textbf{(b)} concerns the super-extremal higher spin particles $|$%
\textsc{m}$_{\mathrm{hs}},$\textsc{q}$_{\mathrm{hs}}>$ emitted by the
unstable higher spin BTZ black holes. These particle states have masses
constrained by \textsc{m}$_{\mathrm{hs}}^{2}\leq $ $M_{BH}^{2};$ see eqs
(3.34-3.35) and (3.75) in \cite{MI}. Additionally, they verify the WGC
condition:
\begin{equation*}
\text{\textsc{m}}_{\mathrm{hs}}^{2}\leq (\frac{1}{l_{\mathrm{AdS}_{3}}}+%
\mathrm{\mu )}^{2}j\left( j+1\right) \leq M_{BH}^{2}
\end{equation*}%
Comparing our conjecture (\ref{M2}) with the general super-extremality
constraint for emitted particles, namely $m^{2}\leq \left( g^{2}/\sqrt{G}%
\right) Q^{2}$ \cite{RWGC}, we identify a correspondence between $\sqrt{%
j\left( j+1\right) }$ and the quantised HS charge \textsc{q}$_{\mathrm{hs}}$%
. Using this correspondence, we obtain \textsc{q}$_{\mathrm{hs}}=\sqrt{%
s\left( s-1\right) }$ where $j=s-1$ as shown in eq(4.10). Moreover, for
sufficiently large enough values of spin $s$, the charge \textsc{q}$_{%
\mathrm{hs}}$\ mirrors the spin in agreement with HS theory labeled by
quantum numbers of sl(2,$\mathbb{R}$) and aligning with the CFT's conserved
currents at the boundary of AdS. Furthermore, this spin-dependent charge
\textsc{q}$_{\mathrm{hs}}$ is not novel in other physical contexts. For
instance, in spintronics, spin accumulation at the boundaries of a material
creates a spin-dependent electric field that links spin to the distribution
of charges. This enables the conversion of a charge current into a spin
current through a process known as the Spin Hall effect \cite{hall} or
conversely through the inverse spin hall effect \cite{hall2}.

\textbf{(i)-} \textbf{Comparison with extremality conditions of Kerr-Newman
type of Black holes}\newline
First, recall that the extremality constraint for Kerr-Newman black hole in
D-dimensions (i.e. \textbf{(a)}-regime black holes) with metric, \textrm{%
\cite{KN}}
\begin{equation*}
ds_{KN}^{2}=-\frac{\mathrm{\delta }}{\rho ^{2}}\left( dt-\mathrm{a}_{{\small %
KN}}\sin ^{2}\theta d\phi \right) ^{2}+\frac{\rho ^{2}}{\mathrm{\delta }}%
dr^{2}+\rho ^{2}d\theta ^{2}+\frac{\sin ^{2}\theta }{\rho ^{2}}\left[
\mathrm{a}_{{\small KN}}dt-\left( r^{2}+\mathrm{a}_{{\small KN}}^{2}\right)
d\phi \right] ^{2}
\end{equation*}%
where $\mathrm{\delta }=r^{2}-2M_{KN}r+\mathrm{a}_{{\small KN}%
}^{2}+Q_{KN}^{2}$ and $\rho =r^{2}+\mathrm{a}_{{\small KN}}^{2}\cos
^{2}\theta ,$ relates the black hole's mass $M_{KN}$, its electric charge $%
Q_{KN}$ and angular momentum $\mathrm{a}_{{\small KN}}$ by the quadratic
relation $M_{KN}^{2}=Q_{KN}^{2}+\mathrm{a}_{{\small KN}}^{2}$ \cite{KN}.
This triangular identity highlights the interchangeable roles of the
electric charge $Q_{KN}^{2}$ and angular momentum $\mathrm{a}_{{\small KN}%
}^{2}$\ in the \emph{extremality} condition ($\mathrm{\delta }=0)$, as the
permutation \ $Q_{KN}\leftrightarrow \mathrm{a}_{{\small KN}}$ leaves the
\emph{extremal} mass $M_{KN}^{2}$ invariant. Since the WGC bounds in the
\textbf{(b)}-regime reflects contributions from the same physical parameters
as in the \textbf{(a)}-regime, we expect an analogous triangular structure
to emerge for the \textbf{(b)}-regime states, as conjectured for the emitted
higher spin particles \textsc{m}$_{\mathrm{hs}}$. In our case, the HS-BTZ
black hole possesses a HS charge and momentum, both of which are similarly
encoded in \textsc{m}$_{\mathrm{hs}}.$\newline
In this regard, we reference a WGC bound that has been proposed in \cite{KNA}
for Kerr-Newman AdS black holes. It remarkably incorporates the extremal
black hole's parameters along with the charge $Q$ of particles in the
vicinity of the black hole horizon. It is of the form:%
\begin{equation*}
\frac{Q{a}_{KN}\left( Q_{KN}\right) _{ext}}{\left( M_{KN}\right)
_{ext}+4a_{KN}^{3}/l^{2}}\geq 1
\end{equation*}%
For particles with charge $Q=\frac{1}{a_{KN}}$ and $\frac{a_{KN}}{l}<<1,$
the bound leads to the extremality constraint $\left( Q_{KN}/M_{KN}\right)
_{ext}\sim O\left( 1\right) .$ This suggests that imposing conditions on
\textbf{(b)}-regimes particles can, a priori, imply constraints on \textbf{%
(a)}-regime black holes. However, can a similar analysis be extended to
HS-BTZ black holes in AdS$_{3}$?

\textbf{(ii)- Comparing (\ref{M2}) with an extremality constraint on HS-BTZ
entropy:}\newline
We begin by recalling that for higher spin BTZ black holes in AdS$_{3},$ an
extremality constraint was proposed for higher spin sl(3) gravity in \cite%
{14}, see also \cite{156, 152}. It is based on deriving entropy using
holonomies of flat connections in Chern-Simons theory of the higher spin
gravity, and then imposing the reality condition, which requires the
inequalities:%
\begin{equation}
|\mathcal{W}_{+}|\leq |\eta \left( \mathcal{L}_{+}\right) ^{3/2}|\qquad
,\qquad |\mathcal{W}_{-}|\leq |\eta \left( \mathcal{L}_{-}\right) ^{3/2}|
\label{I}
\end{equation}%
where ($\mathcal{L}_{+},\mathcal{W}_{+})$ as well as ($\mathcal{L}_{-},%
\mathcal{W}_{-})$\ define the left and \textrm{the} right conserved currents
at the AdS$_{3}$ boundary respectively. To precisely explore the connection
between this extremality condition on the (\textbf{a})-regime black holes
and our (\textbf{b})-regime constraint on higher spin massive particles
within the framework of higher spin topologically \emph{massive} gravity
(HSTMG), we need:\newline
(1)- A constraint relation analogous to the \textrm{reality condition} of
entropy (\ref{I}). Unfortunately, to the best of our knowledge, extremality
constraints for black holes in 3D higher spin massive gravity have yet to be
established.\newline
(2)- To formulate an extremality bound for HS BTZ black holes in HSTMG.
First, one must consider theories beyond the standard massive gravity \cite%
{086} used in our investigation. For instance, minimal models \cite{151}, or
chiral theories \cite{085} could provide a much simplified framework for
entropy calculations. Then, extend these models to incorporate higher spin
fields to compute the entropy in terms of the CS flat connections, impose
the reality condition, deduce the extremality bounds and subsequently
investigate the implications for the weak gravity conjecture. \newline
Given these challenges, we plan to pursue this as a direction for future
research.

\subsection{Refining eqs(\protect\ref{jj}-\protect\ref{cc}) and the
super-extremal tower}

The emergence of quantum numbers of the SL(2,$\mathbb{R}$) representations
in the conjectured mass formula (\ref{M2}) makes one ponder about other
hidden facets of \textsc{m}$_{\mathrm{hs}}^{2}$. Below, we give two
interesting features allowing to refine the eqs(\ref{jj}-\ref{cc}):

The first feature concerns the algebraic interpretation of the expression
\textsc{m}$_{\mathrm{hs}}^{2}$ (\ref{M2}); in fact, \textsc{m}$_{\mathrm{hs}%
}^{2}$ can be perceived as the eigenvalue of a mass operator \textsc{\^{M}}$%
^{2}$ acting on the quantum particle states $\left\vert \Delta
,N\right\rangle $ emitted by the HS-BTZ black hole as follows%
\begin{equation}
\text{\textsc{\^{M}}}^{2}\left\vert \Delta ,N\right\rangle =\text{\textsc{m}}%
_{\mathrm{hs}}^{2}\left\vert \Delta ,N\right\rangle
\end{equation}%
with positive inetegers\textrm{\ }$\Delta $ and N. Acting on these quantum
states $\left\vert \Delta ,N\right\rangle $ by the mass operator%
\begin{equation}
\text{\textsc{\^{M}}}^{2}=\left( M_{\mathrm{AdS}_{3}}+\mathrm{\mu }\right)
^{2}\mathcal{C}_{2}+\left( M_{\mathrm{AdS}_{3}}^{2}-\mathrm{\mu }^{2}\right)
L_{0}  \label{mop}
\end{equation}%
we get its eigenvalues in terms of the quantum numbers $\Delta $ and $N$;
they read as follows
\begin{equation}
\text{\textsc{m}}_{\Delta ,N}^{2}=\left( M_{\mathrm{AdS}_{3}}+\mathrm{\mu }%
\right) ^{2}\Delta \left( \Delta -1\right) +\left( M_{\mathrm{AdS}_{3}}^{2}-%
\mathrm{\mu }^{2}\right) \left( \Delta +N\right)  \label{rhsc}
\end{equation}%
Because $N\in \mathbb{N}$, the quantum states $\left\vert \Delta
,N\right\rangle $ define an infinite tower of states candidates for the
emitted particles of the HS-BTZ black hole. In this regard, recall that in
AdS$_{\mathrm{3}}$\ one must upgrade the mild WGC to stronger forms like the
lattice WGC of \cite{wgc3}. The refined HSC is given by the tower WGC \cite%
{twgc} occupied by super extremal higher spin states (\ref{rhsc}) fulfilling
the mass to charge constraint (\ref{QS}). In fact, the self-interacting
particle condensate stills forms for HSTMG because of the AdS$_{3}$ boundary
conditions, which can act as a box that reflects back the emitted particles,
enabling them to self-interact in a sub-extremal cloud. Our setting and the
physics therein, with the additional mass deformation, is still governed by
the choice of the boundary conditions, similar to the standard theory. This
suggests that the same arguments presented in the standard case \cite{wgc3}
for the WGC refinement beyond the mild version are still applicable.\newline
However, there is another way to justify the need for a refined version for
the class of theories we are considering. Under diagonal boundary
conditions, one can identify the higher spin symmetry $SL(N)\times SL(N)$,
with the affine $U(1)^{2\left( N-1\right) }$ asymptotically. This shows the
presence of multiple U(1) gauge fields at the boundary which further
motivates the need for a WGC formulation beyond the mild version involving a
single U(1) \cite{Soft1, Soft2, BF}.\newline
Now which version of the refined WGC should we apply? Usually, in the
presence of multiple U(1)s, one can impose the convex hall condition \cite%
{CHC}. However, it only requires the emission of a super-extremal vector (a
multi-particle state) which would be problematic for the condensate in this
case. Our proposed refinement is more natural as it is based on the sl(2,$%
\mathbb{R}$) representation. Since the emitted super-extremal states with
masses $($\textsc{m}$_{-}^{2})_{\Delta ,N}$ are closely related to the
unitary SL$_{2}$ representation $\mathcal{R}_{\Delta }^{-},$ the tower of
states fulfilling HS Swampland conjecture was then given by the quantum
states of $\mathcal{R}_{\Delta }^{-}$. We therefore identified the
refinement of the WGC as the tower WGC.

The second feature regards the HS Swampland conjecture (\ref{34D}) namely
\textsc{m}$_{\mathrm{hs}}\leq \sqrt{2}$\textsc{q}$_{\mathrm{hs}}{\large g}_{%
\mathrm{hs}}M_{\mathrm{Pl}}$. This inequality puts a constraint on the
appropriate unitary representation of SL$_{2}$ where the tower of super
extremal particle states $\left\vert \Delta ,N\right\rangle $ emitted by the
HS-BTZ black hole resides. Because $M_{\mathrm{AdS}_{3}}^{2}-\mathrm{\mu }%
^{2}$ has an indefinite sign, we can distinguish three types of mass
operators according to the value of $\mathrm{\mu }^{2}$ compared to $M_{%
\mathrm{AdS}_{3}}^{2}.$ We have%
\begin{equation}
\begin{tabular}{lllll}
$\left( a\right) $ & $:$ & $\mathrm{\mu }^{2}$ & $>$ & $M_{\mathrm{AdS}%
_{3}}^{2}$ \\
$\left( b\right) $ & $:$ & $\mathrm{\mu }^{2}$ & $=$ & $M_{\mathrm{AdS}%
_{3}}^{2}$ \\
$\left( c\right) $ & $:$ & $\mathrm{\mu }^{2}$ & $<$ & $M_{\mathrm{AdS}%
_{3}}^{2}$%
\end{tabular}
\label{regimes}
\end{equation}%
these three phases are very common in the study of TMG theories \cite{tmg1,
tmg2, G, S}. In fact by considering the additional gravitational CS term (%
\ref{gama}), the massive HS gravity theory develops a diffeomorphism anomaly
given by the difference between the right $c_{+}$ and the left $c_{-}$
central charges
\begin{equation}
c_{\pm }=\frac{3l_{\mathrm{AdS}_{{\small 3}}}}{G_{N}}\left( 1\pm \frac{1}{%
\mathrm{\mu }l_{\mathrm{AdS}_{{\small 3}}}}\right)  \label{ch}
\end{equation}%
leading to%
\begin{equation}
\frac{1}{\mathrm{\mu }}=\frac{G_{N}}{6}\left( c_{+}-c_{-}\right)
\end{equation}%
The value of $\mathrm{\mu }\ $is therefore a measure of the violation of
parity in TMG. Additionally, one must note that the central charges are
positive definite when $\frac{1}{\mathrm{\mu }l_{\mathrm{AdS}_{{\small 3}}}}%
\leq 1.$\ As for the critical value $\mu l_{\mathrm{AdS}_{{\small 3}}}=1,$\
it implies the vanishing of the central charges $c_{-}$\ and the resulting
TMG theory was shown to be dual to a logarithmic CFT \cite{22}.

For all three phases (\ref{regimes}), the mass operator takes the following
forms%
\begin{equation}
\begin{tabular}{lllll}
$\left( a\right) $ & $:$ & $\text{\textsc{\^{M}}}_{-}^{2}$ & $=$ & $\left(
M_{\mathrm{AdS}_{3}}+\mathrm{\mu }\right) ^{2}\mathcal{C}_{2}-\left\vert M_{%
\mathrm{AdS}_{3}}^{2}-\mathrm{\mu }^{2}\right\vert L_{0}$ \\
$\left( b\right) $ & $:$ & $\text{\textsc{\^{M}}}_{0}^{2}$ & $=$ & $4M_{%
\mathrm{AdS}_{3}}^{2}\mathcal{C}_{2}$ \\
$\left( c\right) $ & $:$ & $\text{\textsc{\^{M}}}_{+}^{2}$ & $=$ & $\left(
M_{\mathrm{AdS}_{3}}+\mathrm{\mu }\right) ^{2}\mathcal{C}_{2}+\left\vert M_{%
\mathrm{AdS}_{3}}^{2}-\mathrm{\mu }^{2}\right\vert L_{0}$%
\end{tabular}%
\end{equation}%
Acting by these operators on the particle states $\left\vert \Delta
,N\right\rangle ,$ we obtain the eigenvalues%
\begin{equation}
\begin{tabular}{lll}
(\textsc{m}$_{-}^{2}$)$_{\Delta ,N}$ & $=$ & $\left( M_{\mathrm{AdS}_{3}}+%
\mathrm{\mu }\right) ^{2}\Delta \left( \Delta -1\right) -\left\vert M_{%
\mathrm{AdS}_{3}}^{2}-\mathrm{\mu }^{2}\right\vert \left( \Delta +N\right) $
\\
(\textsc{m}$_{0}^{2}$)$_{\Delta ,N}$ & $=$ & $4M_{\mathrm{AdS}%
_{3}}^{2}\Delta \left( \Delta -1\right) $ \\
(\textsc{m}$_{+}^{2}$)$_{\Delta ,N}$ & $=$ & $\left( M_{\mathrm{AdS}_{3}}+%
\mathrm{\mu }\right) ^{2}\Delta \left( \Delta -1\right) +\left\vert M_{%
\mathrm{AdS}_{3}}^{2}-\mathrm{\mu }^{2}\right\vert \left( \Delta +N\right) $%
\end{tabular}
\label{pm}
\end{equation}%
which for $\Delta >1,$ they obey the inequalities%
\begin{equation}
\begin{tabular}{lll}
(\textsc{m}$_{-}^{2}$)$_{\Delta ,N}$ & $<$ & $\left( M_{\mathrm{AdS}_{3}}+%
\mathrm{\mu }\right) ^{2}\Delta \left( \Delta -1\right) $ \\
(\textsc{m}$_{0}^{2}$)$_{\Delta ,N}$ & $=$ & $4M_{\mathrm{AdS}%
_{3}}^{2}\Delta \left( \Delta -1\right) $ \\
(\textsc{m}$_{+}^{2}$)$_{\Delta ,N}$ & $>$ & $\left( M_{\mathrm{AdS}_{3}}+%
\mathrm{\mu }\right) ^{2}\Delta \left( \Delta -1\right) $%
\end{tabular}%
\end{equation}%
showing that (\textsc{m}$_{0}^{2}$)$_{\Delta ,N}$ is a critical mass. This
feature allows to think about the (\textsc{m}$_{-}^{2}$)$_{\Delta ,N}$
inequality as follows
\begin{equation}
(\text{\textsc{m}}_{-}^{2})_{\Delta ,N}\leq 4M_{\mathrm{AdS}_{3}}^{2}\Delta
\left( \Delta -1\right) \qquad \Leftrightarrow \qquad \text{\textsc{m}}_{%
\mathrm{hs}}\leq \sqrt{2}\text{\textsc{q}}_{\mathrm{hs}}{\large g}_{\mathrm{%
hs}}M_{\mathrm{Pl}}  \label{r21}
\end{equation}%
from which we deduce the HS charge \textsc{q}$_{\mathrm{hs}}$ and the HS
coupling constant ${\large g}_{\mathrm{hs}}$ supported by the representation
theory,
\begin{equation}
\text{\textsc{q}}_{\mathrm{hs}}=\sqrt{\Delta \left( \Delta -1\right) }\qquad
\Leftrightarrow \qquad {\large g}_{\mathrm{hs}}=\sqrt{2}\frac{M_{\mathrm{AdS}%
_{3}}}{M_{\mathrm{Pl}}}
\end{equation}%
Notice finally that expressing eqs(\ref{pm}) as%
\begin{equation}
(\text{\textsc{m}}_{\pm }^{2})_{\Delta ,N}=\left( M_{\mathrm{AdS}_{3}}+%
\mathrm{\mu }\right) ^{2}\Delta \left( \Delta -1\right) \pm \left\vert M_{%
\mathrm{AdS}_{3}}^{2}-\mathrm{\mu }^{2}\right\vert \left( \Delta +N\right)
\label{tower}
\end{equation}%
we see that these masses $($\textsc{m}$_{\pm }^{2})_{\Delta ,N}$ are
intimately related to the unitary SL$_{2}$ representations $\mathcal{R}%
_{\Delta }^{\pm }.$ The tower of states fulfilling HS Swampland conjecture (%
\ref{r21}) is then given by the quantum states of $\mathcal{R}_{\Delta }^{-}$%
.

\section{Piecing HSC in the WGC framework}

\label{sec:5} The weak gravity conjecture is one of the seminal ideas in the
swampland program, and may very well be the most properly argued swampland
criteria. It has been studied in numerous settings with various
parametrisations and configurations, giving many formulations that differ
both in their assumptions as well as in their regime of applicability, for
an extensive review refer to \cite{rev}. Pertaining to our concern, we will
briefly look over some of its statements and implications for AdS theories.

\subsection{WGC in AdS background}

A prerequisite of any potential WGC formulation in a curved AdS$_{d}$ space
is the possibility to recover the usual bound of the flat space once the
curvature $l_{AdS_{\mathrm{d}}}\rightarrow \infty $ \cite{rev}.
Unfortunately, a general AdS formulation of the WGC is still a pending
issue. However, there are many proposals like the one in \cite{53}:%
\begin{equation}
\frac{\delta ^{2}}{l_{AdS_{\mathrm{d}}}^{2}}\leq \frac{d-2}{d-3}\frac{%
e^{2}q^{2}}{G_{N}^{2}}  \label{53}
\end{equation}%
where $\delta $ is the conformal scaling dimension related to the mass $%
\mathbf{m}$ via%
\begin{equation}
\delta =\frac{d-1}{2}+\sqrt{\frac{\left( d-1\right) ^{2}}{4}+l_{AdS_{\mathrm{%
d}}}^{2}\mathbf{m}^{2}}
\end{equation}%
In addition to the bound (\ref{53}) having the d=3 singularity, it is not
satisfied for all CFTs and it is unclear why this particular condition is
most likely to hold universally \cite{rev}. Another Anti de Sitter WGC
reformulation is given by the charge convexity conjecture \cite{3C}, it
imposes bounds in terms of binding energy using the lowest dimension
operator of the associated CFT$.$ Although the convex charge constraint is
believed to be more general than the WGC, we disregard it as it differs from
the usual statements motivated by black holes decay or long range forces.

To overcome the triviality of the constraint (\ref{53}) for d=3, there is an
alternative method that exploits tools of the AdS$_{3}$/CFT$_{2}$
correspondence. In \textrm{\cite{wgc3}} and more generally in \textrm{\cite%
{wwgc3}, }the weak gravity conjecture was indeed derived using a conformal
approach by demanding the partition function of the boundary CFT$_{2}$ to be
modular invariant. In a disjointed setting \textrm{\cite{wwgc3},} where the
gravitational and gauge sectors are distinct by considering 3D gravity in
addition to a U(1) gauge field, it is possible to establish a constraint on
the conformal dimension of the lightest charged state as follows \textrm{%
\cite{wwgc3},}%
\begin{equation}
\delta -\delta _{\text{\textsc{vac}}}\simeq \frac{c}{6}+\frac{3}{2\pi }+%
\mathcal{O}\left( \frac{1}{c}\right)  \label{wgc}
\end{equation}%
This bound is not optimal, and can be enhanced via additional symmetries. In
fact, for 2D supersymmetric CFT with $\mathcal{N}=(1,1)$ supercharges, the
constraint (\ref{wgc}) on the conformal weight improves to $\delta \simeq
1+O\left( 1/c\right) .$\newline
However, the constraint (\ref{wgc}) isn't suitable for HS-TMG as it doesn't
consider charged higher spin fields and only concerns U(1) charges.%
\begin{equation}
\begin{tabular}{|c|c|c|c|c|c|}
\hline
{\small Formulations} &
\begin{tabular}{c}
{\small AdS} \\
{\small background}%
\end{tabular}
& {\small D=3} &
\begin{tabular}{c}
{\small BH} \\
{\small solution}%
\end{tabular}
&
\begin{tabular}{c}
{\small Massive} \\
{\small HS fields}%
\end{tabular}
& {\small HS charge} \\ \hline
{\small WGC in AdS \cite{53}} & {\small x} & {\small -} & {\small x} &
{\small -} & {\small -} \\ \hline
\begin{tabular}{c}
{\small Convex Charge} \\
{\small Constraint \cite{3C}}%
\end{tabular}
& {\small x} & {\small x} & {\small -} & {\small -} & {\small -} \\ \hline
{\small WGC in AdS}$_{3}${\small \ \cite{wgc3}} & {\small x} & {\small x} &
{\small x} & {\small -} & {\small -} \\ \hline
\end{tabular}%
\end{equation}
\newline

\subsection{Beyond electric U(1) charges}

There are other formulations of the WGC that experimented with parameters
beyond the typical electric U(1) charges. For instance, the so called
spinning weak gravity conjecture \cite{spinwgc} where\textrm{\ }quantum or
higher derivative corrections lead to perturbed (BTZ) black holes obeying a
rotating version of the WGC that follows from the holographic c-theorem.
Another interesting case is the causality bounds on higher spin particles
coupled to stringy gravity in 4D \cite{hs}. In fact, in order for a 4D
gravitational theory coupled to a tower of higher spin states to be causal,
a WGC-like constraint must be imposed on the lightest HS particle. The 4D
causality bound is reminiscent of the spin-2 conjecture requiring a cutoff
on gravitational theories with massive higher spin fields \cite{spin2}.%
\begin{equation}
\begin{tabular}{|c|c|c|c|c|c|}
\hline
{\small Formulations} &
\begin{tabular}{c}
{\small AdS} \\
{\small background}%
\end{tabular}
& {\small D=3} &
\begin{tabular}{c}
{\small BH} \\
{\small solution}%
\end{tabular}
&
\begin{tabular}{c}
{\small Massive} \\
{\small HS fields}%
\end{tabular}
& {\small HS charge} \\ \hline\hline
{\small WGC in AdS \cite{53}} & {\small x} & {\small -} & {\small x} &
{\small -} & {\small -} \\ \hline
\begin{tabular}{c}
{\small Convex Charge} \\
{\small Constraint \cite{3C}}%
\end{tabular}
& {\small x} & {\small x} & {\small -} & {\small -} & {\small -} \\ \hline
{\small WGC in AdS}$_{3}${\small \ \cite{wgc3}} & {\small x} & {\small x} &
{\small x} & {\small -} & {\small -} \\ \hline
\begin{tabular}{c}
{\small A spinning} \\
{\small WGC \cite{spinwgc}}%
\end{tabular}
& {\small x} & {\small x} & {\small x} & {\small -} & {\small -} \\ \hline
{\small HS causality \cite{hs}} & {\small -} & {\small -} & {\small x} &
{\small x} & {\small -} \\ \hline
{\small Our HSC proposal} & {\small x} & {\small x} & {\small x} & {\small x}
& {\small x} \\ \hline\hline
\end{tabular}%
\end{equation}%
\begin{equation*}
\end{equation*}%
As evidenced, the HSC addresses a setting with a particular configuration to
investigate the WGC. We derive a WGC-like constraint for black hole
solutions of higher spin topological massive gravity carrying higher spin
charges. The HSC stems from the core SL(2) algebraic representations and
provides a constraint on the HS fields masses and charges to regulate the
discharge of the HS BTZ solutions. Before further discussing the difference
between the HSTMG and the more standard setup of the WGC with local U(1)
degrees of freedom, let us review some of the main similarities. \newline
For a higher spin gravity theory with $SL(N)\times SL(N)$ symmetry, imposing
diagonal boundary conditions generates asymptotic symmetries governed by the
$U(1)^{\left( N-1\right) }\times U(1)^{\left( N-1\right) }$ affine algebra
\cite{Soft1, Soft2, BF}. The higher spin particles, higher spin versions of
the graviton, emerge through composites of the $U(1)$ photons via a twisted
Sugawara construction at the boundary. The BTZ black hole solution in this $%
SL(N)$ higher spin gravity theory is therefore analogous to a charged BTZ
black hole solution in AdS$_{3}$ Einstein gravity with SL(2)$\times $SL(2)
coupled to $U(1)^{\left( N-2\right) }\times U(1)^{\left( N-2\right) }$ gauge
fields$. $ In this case, we introduced massless higher spin degrees of
freedom, endowing the BTZ black hole with higher spin charges which
correspond to $U(1)$ charges in the diagonal representation.\newline
However, with the inclusion of the gravitational Chern-Simons term, we
induce a mass deformation in the theory's geometry. This is evident from the
modified equations of motion $G_{\mu \nu }+\frac{1}{\mathrm{\mu }}C_{\mu \nu
}=0$ having non vanishing Cotton tensor $C_{\mu \nu }\neq 0$ due to the
presence of the CS gravitational term. Therefore, unlike the additional
gauge charges, the CS gravitational term invokes a mass deformation,
yielding massive higher spin degrees of freedom that effect the geometry of
the spacetimes and the associated metrics. It becomes necessary to adapt the
WGC to these new degrees of freedom that affect the black hole's stability
and dynamics.\newline
Exploring swampland conjectures from the lens of holographic theories has
been of great interest recently. While we mainly focused on the WGC, there
is a substantial body of work relating the swampland distance conjecture to
higher spin theories as in \cite{6} and the ensuing \cite{7, 8}. For
instance in \cite{6}, it has been proposed that at infinite distances all
theories possess an emergent HS symmetry in such a manner that certain
proprieties of the conformal manifolds can be written as a function of the
HS spectrum.

\section{Conclusion}

\label{sec:6} In this paper, we investigated a well motivated inquiry
regarding the discharge of higher spin BTZ black holes in a higher spin
topological massive gravity setting with Chern-Simons formulation based on
rank-2 higher spin gauge symmetries. We proposed a higher spin Swampland
conjecture to regulate the emission of super-extremal higher spin particles
given by an upper bound on their mass to charge ratio.

En route to derive the higher spin swampland conjecture, we first
established a correspondence between the massive higher spin AdS$_{3}$
models and effective gauge theories coupled to D-gravity (EFF$_{{\small D}}$%
) to hypothesize a formulation of the swampland constraint for higher spin
BTZ black holes. Exploiting the principal SL(2,$\mathbb{R}$) of the higher
spin gauge symmetry, we constructed the charge (\ref{QHS}) and the mass (\ref%
{mop}) operators as well as their eigenvalues (\ref{pm}, \ref{jj}). We also
computed the higher spin gauge coupling constant (\ref{cc}) and showcased
its relation to the inverse of the Chern-Simons level \textrm{k (\ref{cc2})}.

Furthermore by using the infinite dimensional unitary representations,
particularly the discrete series $\mathcal{R}_{\Delta }^{-},$ we built a
tower of higher spin states (\ref{tower}) occupied by the emitted higher
spin particles in accordance with the lattice refinement required for the AdS%
$_{3}$ space. We must note that the mass operator leading to the tower of
higher spin states ensues from the phase $\mathrm{\mu }^{2}>M_{\mathrm{AdS}%
_{3}}^{2}$ assuring the positivity of the central charges (\ref{ch}) as well
as the unitarity of the CFT.

On a final note, we discussed the various WGC formulations especially for
AdS backgrounds in different settings to place the higher spin swampland
conjecture within the WGC framework as a way to emphasize the pertinence of
our work regarding recent advancements in the swampland program. Overall,
the antecedent results may imply several interpretations:

\begin{description}
\item[$(\mathbf{i})$] The inclusivity of topological massive gravity within
the general Landscape of consistent quantum gravitational theories.

\item[$(\mathbf{ii})$] Particularly, the established link between the higher
spin conjecture and the WGC constraint conveys the validity of the later for
topological massive higher spin gravitational models.

\item[$(\mathbf{iii})$] The existence of the tower of higher spin states is
strongly supported by algebraic properties of the core SL(2,$\mathbb{R}$) of
the HS gravity namely the discrete infinite unitary representation $\mathcal{%
R}_{\Delta }^{-}$.%
\begin{equation*}
\end{equation*}
\end{description}


\begin{thebibliography}{99}
\bibitem{swp1} Vafa, C. (2005). The String landscape and the swampland.
arXiv preprint hep-th/0509212.

\bibitem{palti} Palti, E. (2019). The swampland: introduction and review.
Fortschritte der Physik, 67(6), 1900037.

\bibitem{val} van Beest, M., Calderon Infante, J., Mirfendereski, D., and
Valenzuela, I. (2022). Lectures on the swampland program in string
compactifications. Physics Reports, 989, 1-50.

\bibitem{alvarez} Alvarez Garcia, R., Blumenhagen, R., Kneissl, C.,
Makridou, A., and Schlechter, L. (2022). Swampland conjectures for an almost
topological gravity theory. Physics Letters B, 825, 136861.

\bibitem{holo} Ashwinkumar, M., Leedom, J. M., and Yamazaki, M. (2023).
Duality Origami: Emergent Ensemble Symmetries in Holography and Swampland.
arXiv preprint arXiv:2305.10224.

\bibitem{tmg1} Deser, S., Jackiw, R., and Templeton, S. (1982).
Three-dimensional massive gauge theories. Physical Review Letters, 48(15),
975.

\bibitem{tmg2} Deser, S., Jackiw, R., and Templeton, S. (2000).
Topologically massive gauge theories. Annals of Physics, 281(1-2), 409-449.

\bibitem{hstmg1} Chen, B., and Long, J. (2011). High spin topologically
massive gravity. Journal of High Energy Physics, 2011(12), 1-14.

\bibitem{hstmg2} Kuzenko, S. M., and Ponds, M. (2018). Topologically massive
higher spin gauge theories. Journal of High Energy Physics, 2018(10), 1-44.

\bibitem{BTZ} Banados, M., Teitelboim, C., and Zanelli, J. Black hole in
three-dimensional spacetime. Physical Review Letters, 69(13), 1849, (1992).

\bibitem{BTZ1} Li, W., Song, W., and Strominger, A. (2008). Chiral gravity
in three dimensions. Journal of High Energy Physics, 2008(04), 082.

\bibitem{nonsusy} Ooguri, H., and Vafa, C. (2016). Non-supersymmetric AdS
and the Swampland. arXiv preprint arXiv:1610.01533.

\bibitem{nonsysy2} Freivogel, B., and Kleban, M. (2016). Vacua morghulis.
arXiv preprint arXiv:1610.04564.

\bibitem{wgc} Arkani-Hamed, N., Motl, L., Nicolis, A., and Vafa, C. (2007).
The String landscape, black holes and gravity as the weakest force. Journal
of High Energy Physics, 2007(06), 060.

\bibitem{raja} Sammani, R., Boujakhrout, Y., Laamara, R. A., and Drissi, L.
B. (2024). Finiteness of 3D higher spin gravity Landscape. Classical and
Quantum Gravity, 41(21), 215012.

\bibitem{charkaoui} Charkaoui, M., Sammani, R., Saidi, E. H., and Laamara,
R. A. (2024). Asymptotic Weak Gravity Conjecture in M-theory on K 3\~{A}---
K 3. Progress of Theoretical and Experimental Physics, 2024(7), 073B08.

\bibitem{wgc3} Montero, M., Shiu, G., and Soler, P. (2016). The weak gravity
conjecture in three dimensions. Journal of High Energy Physics, 2016(10),
1-36.

\bibitem{wwgc3} Benjamin, N., Dyer, E., Fitzpatrick, A. L., and Kachru, S.
(2016). Universal bounds on charged states in 2d CFT and 3d gravity. Journal
of High Energy Physics, 2016(8), 1-26.

\bibitem{AT} Achucarro, A., and Townsend, P. K. A Chern-Simons action for
three-dimensional anti-de Sitter supergravity theories. Physics Letters B,
180 (1-2), 89-92. (1986).

\bibitem{W} Witten, E. $2+1$ dimensional gravity as an exactly soluble
system. Nuclear Physics B, 311(1), 46-78. (1988).

\bibitem{spin3} Gutperle, M., and Kraus, P. (2011). Higher spin black holes.
Journal of High Energy Physics, 2011(5), 1-24.

\bibitem{slN} Kraus, P., and Perlmutter, E. (2011). Partition functions of
higher spin black holes and their CFT duals. Journal of High Energy Physics,
2011(11), 1-25.

\bibitem{rajae} Sammani, R., Boujakhrout, Y., Saidi, E. H., Laamara, R. A.,
and Drissi, L. B. (2023). Higher spin AdS 3 gravity and Tits-Satake
diagrams. Physical Review D, 108(10), 106019.

\bibitem{trunc} Chen, B., Long, J., and Wang, Y. (2012). Black holes in
truncated higher spin AdS3 gravity. Journal of High Energy Physics,
2012(12), 1-21.

\bibitem{Cab} Eberhardt, L., Gaberdiel, M. R., and Gopakumar, R. (2020).
Deriving the AdS3/CFT2 correspondence. Journal of High Energy Physics,
2020(2), 1-52.

\bibitem{sahraoui} Sahraoui, E.M., and Saidi, E.H. (2003) Metric Building of
pp Wave Orbifold Geometries, Phys.Lett. B558 221-228, arXiv:hep-th/0210168.

\bibitem{Goddard} Goddard, P. (1986). Vertex operators and algebras.
Superstrings, Supergravity and Unified Theories, 255-291.

\bibitem{BH} Brown, J. D., and Henneaux, M. (1986). Central charges in the
canonical realization of asymptotic symmetries: An example from three
dimensional gravity. Communications in Mathematical Physics, 104, 207-226.

\bibitem{BAR} Bargmann, V. (1947). Irreducible unitary representations of
the Lorentz group. Annals of Mathematics, 48(3), 568-640.

\bibitem{BEN} Benkaddour, I., Rhalami, A. E., and Saidi, E. H. (2001).
Non-trivial extension of the (1+ 2)-Poincar\~{A}\copyright\ algebra and
conformal invariance on the boundary of. The European Physical Journal
C-Particles and Fields, 21(4), 735-747.

\bibitem{BE} Traubenberg, M. R. D., and Slupinski, M. J. (1997). Nontrivial
extensions of the 3D-Poincar\~{A}\copyright\ algebra and fractional
supersymmetry for anyons. Modern Physics Letters A, 12(39), 3051-3066.

\bibitem{GR} Grumiller, D.,and Riegler, M.. Most general AdS3 boundary
conditions. JHEP, 2016(10), 1-23. (2016).

\bibitem{castro} Castro, A. (2016). Lectures on higher spin black holes in
AdS3 gravity. Acta Phys. Polon. B, 47, 2479.

\bibitem{EHS} Saidi, E. H. (2020). Quantum line operators from Lax pairs.
Journal of Mathematical Physics, 61(6).

\bibitem{YEH} Boujakhrout, Y. (2022). On exceptional't Hooft lines in
4D-Chern-Simons theory. Nuclear Physics B, 980, 115795.

\bibitem{compleoni} Campoleoni, A., Fredenhagen, S., and Pfenninger, S.
(2011). Asymptotic $\mathcal{W} $-symmetries in three-dimensional
higher-spin gauge theories. Journal of High Energy Physics, 2011(9), 1-57.

\bibitem{22} Carlip, S., Deser, S., Waldron, A., and Wise, D. K. (2009).
Cosmological topologically massive gravitons and photonsSome of these
results were reported in abbreviated form in [1].

\bibitem{23} Carlip, S., Deser, S., Waldron, A., and Wise, D. K. (2008).
Topologically massive AdS gravity. Physics Letters B, 666(3), 272-276.

\bibitem{mass} Aharony, O., Gubser, S. S., Maldacena, J., Ooguri, H., and
Oz, Y. (2000). Large N field theories, string theory and gravity. Physics
Reports, 323(3-4), 183-386.

\bibitem{mass12} Penedones, J. (2017). TASI lectures on AdS/CFT. In New
Frontiers in Fields and Strings: TASI 2015 Proceedings of the 2015
Theoretical Advanced Study Institute in Elementary Particle Physics (pp.
75-136).

\bibitem{mass13} Gaberdiel, M. R., and Gopakumar, R. (2011). An AdS3 dual
for minimal model CFTs. Physical Review D---Particles, Fields, Gravitation,
and Cosmology, 83(6), 066007.

\bibitem{mass2} Gobeil, Y. (2017). Casimirs of the conformal group. McGill
University (Canada).

\bibitem{mass3} R\"{u}hl, W. (2005). The masses of gauge fields in higher
spin field theory on AdS (4). Physics Letters B, 605(3-4), 413-418.

\bibitem{mass4} Zaffaroni, A. (2000). Introduction to the AdS-CFT
correspondence. Classical and Quantum Gravity, 17(17), 3571.

\bibitem{RWGC} Harlow, D., Heidenreich, B., Reece, M., \& Rudelius, T.
(2022). The weak gravity conjecture: a review. arXiv preprint
arXiv:2201.08380.

\bibitem{MI} Heidenreich, B., \& Lotito, M. (2024). Proving the Weak Gravity
Conjecture in Perturbative String Theory, Part I: The Bosonic String. arXiv
preprint arXiv:2401.14449.

\bibitem{KN} Hod, S. (2015). Extremal Kerr--Newman black holes with
extremely short charged scalar hair. Physics Letters B, 751, 177-183.

\bibitem{KNA} Sadeghi, J., Shokri, M., Alipour, M. R., \& Gashti, S. N.
(2023). Weak gravity conjecture from conformal field theory: a challenge
from hyperscaling violating and Kerr-Newman-AdS black holes. Chinese Physics
C, 47(1), 015103.

\bibitem{14} Bunster, C., Henneaux, M., Perez, A., Tempo, D., \& Troncoso,
R. (2014). Generalized black holes in three-dimensional spacetime. Journal
of High Energy Physics, 2014(5), 1-55.

\bibitem{156} Henneaux, M., P\'{e}rez, A., Tempo, D., \& Troncoso, R.
(2015). Hypersymmetry bounds and three-dimensional higher-spin black holes.
Journal of High Energy Physics, 2015(8), 1-25.

\bibitem{152} Ba\~{n}ados, M., Castro, A., Faraggi, A., \& Jottar, J. I.
(2016). Extremal higher spin black holes. Journal of High Energy Physics,
2016(4), 1-69.

\bibitem{151} Setare, M. R., \& Adami, H. (2015). Entropy formula of black
holes in minimal massive gravity and its application for BTZ black holes.
Physical Review D, 91(10), 104039.

\bibitem{witten} Witten, E. (1998). Anti de Sitter space and holography.
arXiv preprint hep-th/9802150.

\bibitem{085} Ertl, S., Grumiller, D., \& Johansson, N. (2009). Erratum
toInstability in cosmological topologically massive gravity at the chiral
point', arXiv: 0805.2610. arXiv preprint arXiv:0910.1706.

\bibitem{086} Myung, Y. S., Lee, H. W., \& Kim, Y. W. (2008). Entropy of
black holes in topologically massive gravity. arXiv preprint arXiv:0806.3794.

\bibitem{extr} Hod, S. (2015). Extremal Kerr--Newman black holes with
extremely short charged scalar hair. Physics Letters B, 751, 177-183.

\bibitem{hall} Dyakonov, M. I., and Khaetskii, A. V. (2008). Spin hall
effect. Spin physics in semiconductors, 211-243.

\bibitem{hall2} Ando, K., and Saitoh, E. (2012). Observation of the inverse
spin Hall effect in silicon. Nature communications, 3(1), 629.

\bibitem{twgc} Andriolo, S., Junghans, D., Noumi, T., and Shiu, G. (2018). A
tower weak gravity conjecture from infrared consistency. Fortschritte der
Physik, 66(5), 1800020.

\bibitem{Soft1} Afshar, H., Detournay, S., Grumiller, D., Merbis, W., Perez,
A., Tempo, D., and Troncoso, R. (2016). Soft Heisenberg hair on black holes
in three dimensions. Physical Review D, 93(10), 101503.

\bibitem{Soft2} Grumiller, D., P\~{A}\copyright rez, A., Prohazka, S.,
Tempo, D., and Troncoso, R. (2016). Higher spin black holes with soft hair.
Journal of High Energy Physics, 2016(10), 1-25.

\bibitem{BF} Sammani, R., Saidi, E.H. Black flowers and real forms of higher
spin symmetries. J. High Energ. Phys. 2024, 44 (2024).
https://doi.org/10.1007/JHEP10(2024)044

\bibitem{CHC} Cheung, C., and Remmen, G. N. (2014). Naturalness and the weak
gravity conjecture. Physical review letters, 113(5), 051601.

\bibitem{G} Grumiller, D., and Johansson, N. (2008). Instability in
cosmological topologically massive gravity at the chiral point. Journal of
High Energy Physics, 2008(07), 134.

\bibitem{S} Bagchi, A., Lal, S., Saha, A., and Sahoo, B. (2011).
Topologically massive higher spin gravity. Journal of High Energy Physics,
2011(10), 1-28.

\bibitem{rev} Harlow, D., Heidenreich, B., Reece, M., and Rudelius, T.
(2022). The weak gravity conjecture: a review. arXiv preprint
arXiv:2201.08380.

\bibitem{53} Nakayama, Y., and Nomura, Y. (2015). Weak gravity conjecture in
the AdS/CFT correspondence. Physical Review D, 92(12), 126006.

\bibitem{3C} Aharony, O., and Palti, E. (2021). Convexity of charged
operators in CFTs and the weak gravity conjecture. Physical Review D,
104(12), 126005.

\bibitem{spinwgc} Aalsma, L., Cole, A., Loges, G. J., and Shiu, G. (2021). A
new spin on the weak gravity conjecture. Journal of High Energy Physics,
2021(3), 1-39.

\bibitem{hs} Kaplan, J., and Kundu, S. (2021). Closed strings and weak
gravity from higher-spin causality. Journal of High Energy Physics, 2021(2),
1-44.

\bibitem{spin2} Klaewer, D., Lust, D., and Palti, E. (2019). A Spin-2
Conjecture on the Swampland. Fortschritte der Physik, 67(1-2), 1800102..

\bibitem{6} Perlmutter, E., Rastelli, L., Vafa, C., and Valenzuela, I.
(2021). A CFT distance conjecture. Journal of High Energy Physics, 2021(10),
1-33.

\bibitem{7} Campoleoni, A. (2022, October). Infinite distances in
multicritical CFTs and higher-spin holography. In Higher Spin Gravity and
its Applications.

\bibitem{8} Baume, F., and Calderon Infante, J. (2023). On higher-spin
points and infinite distances in conformal manifolds. Journal of High Energy
Physics, 2023(12), 1-51.
\end{thebibliography}
\end{document}